\documentclass[twocolumn]{aastex631}
\usepackage{amsmath}
\usepackage{amssymb}
\usepackage{isomath}
\newcommand{\bvec}[1]{\hat{\mathbfit{#1}}}

\shorttitle{CorrCal}
\shortauthors{Pascua, Sievers, \& Liu}

\begin{document}

\title{Correlation Calibration: \\ A Hybrid Calibration Technique for Radio Interferometric Arrays}

\correspondingauthor{Robert Pascua}
\email{robert.pascua@utoronto.ca}

\author[0000-0003-0073-5528]{Robert Pascua}
\affiliation{Department of Physics and Trottier Space Institute, McGill University \\
3600 University Street \\
Montreal, QC H3A 2T8, Canada}
\affiliation{Dunlap Institute for Astronomy and Astrophysics, University of Toronto \\
50 St. George Street \\
Toronto, ON M5S 3H4, Canada}
\affiliation{Perimeter Institute for Theoretical Physics \\ 31 Caroline Street North \\ Waterloo, ON N2L 2Y5, Canada}
\author[0000-0001-6903-5074]{Jonathan Sievers}
\affiliation{Department of Physics and Trottier Space Institute, McGill University \\
3600 University Street \\
Montreal, QC H3A 2T8, Canada}
\author[0000-0001-6876-0928]{Adrian Liu}
\affiliation{Department of Physics and Trottier Space Institute, McGill University \\
3600 University Street \\
Montreal, QC H3A 2T8, Canada}

\begin{abstract}
Calibrating out per-antenna signal chain effects is an essential step in analyzing radio interferometric data.
For drift-scanning arrays, robustly calibrating the data is especially challenging due to the lack of the ability to track a calibration source.
Consequently, calibration strategies for drift-scanning arrays are limited by our knowledge of the radio sky at large, as well as the direction-dependent instrument response.
In the context of 21\,cm cosmology, where small calibration errors can conspire to overwhelm the cosmological signal, it is therefore crucially important to develop calibration strategies that are capable of accurately calibrating the data in the presence of sky or instrument modeling errors.
In this paper we present CorrCal, a covariance-based calibration strategy for radio interferometric arrays.
CorrCal is a hybrid calibration strategy that leverages the strengths of traditional sky-based calibration and redundant calibration in a computationally efficient framework that is fairly insensitive to modeling errors.
We find that calibration errors from CorrCal are unbiased and far below typical thermal noise thresholds across a wide range of modeling error scenarios.
We show that CorrCal is computationally efficient: our implementation is capable of evaluating the likelihood and its gradient in less than a second for 1,000-element class arrays using just a single laptop core.
Given CorrCal's computational efficiency and robustness to modeling errors, we anticipate that it will serve as a useful tool in the analysis of radio interferometric data from current and next-generation experiments targeting the cosmological 21\,cm signal.
\end{abstract}

\section{Introduction}
\label{sec:INTRODUCTION}
Calibration is a fundamental challenge in the analysis of radio interferometric data.
Calibration takes the raw interferometric visibilities, which have been modulated by the electronics in the signal chain, and converts them into physically meaningful quantities by establishing an absolute flux scale and a common phase center, enabling downstream analyses such as mapmaking or power spectrum estimation.
Traditional approaches to calibration tend to rely on the ability to steer the individual array elements, since it is fairly straightforward to establish an absolute flux scale and calibrate out the relative phases between antennas when the array is pointing at a bright, well-known source of radio emission that dominates the measured visibilities.
Many current and next-generation arrays, however, have opted to instead take a drift-scan approach, where array elements are not steerable but instead observe different patches of sky in accordance with the Earth's rotation.
Notably, experiments such as the Hydrogen Epoch of Reionization Array~\citep[HERA,][]{DeBoer:2017,Berkhout:2024}, the Canadian Hydrogen Intensity Mapping Experiment~\citep[CHIME,][]{CHIME:2022}, the Canadian Hydrogen Observatory and Radio-transient Detector~\citep[CHORD,][]{Vanderlinde:2020}, and the Hydrogen Intensity and Real-time Analysis eXperiment~\citep[HIRAX,][]{Newburgh:2016,Crichton:2022} have adopted a drift-scan approach because it allows for a more economical array element design that can be produced at scale.
With the increasing prevalence of drift-scanning arrays, there is an increasing need for innovative approaches to calibration---in the absence of a reliable calibration source, it is much more difficult to establish an absolute reference.
These new drift-scanning arrays also tend to operate at low frequencies, where our knowledge of the sky and the instrument response are both highly uncertain.
The need for improved calibration routines is accentuated in the context of 21\,cm cosmology, where the extreme dynamic range between the cosmological signal and the astrophysical foreground signal~\citep{Liu&Shaw:2020} demands exquisite control over systematic effects in the data~\citep[e.g.,][]{Barry:2016,Joseph:2018,Byrne:2019,Orosz:2019}.
In this paper, we present Correlation Calibration, which we will hereafter refer to as \emph{CorrCal}, a hybrid approach to calibration that simultaneously leverages the strengths of calibration routines often employed in drift-scanning arrays and significantly reduces the calibration errors induced by modeling errors.

While there exists a myriad of approaches to calibration~\citep[e.g.,][]{Yatawatta:2009,Liu:2010,Dillon:2018,Byrne:2021,Ewall-Wice:2022b,Sims:2022a,Sims:2022b,Byrne:2023,Cox:2024}, the majority of direction-independent calibration algorithms tend to fall into one of two categories: sky-based calibration or redundant calibration.
In sky-based calibration, the per-antenna direction-independent \emph{gains} are obtained by modeling radio sources on the sky and the direction-dependent instrument response.
The expected visibilities measured by the array are then computed from the sky model and instrument response and the per-antenna gains are adjusted to find the best-fit solution given the observed data.
Naturally, sky-based calibration is prone to errors in modeling the radio sky as well as errors in modeling the instrument response.
In redundant calibration, the expected visibilities are not computed from a sky and instrument model, but are instead treated as free parameters that are fit for in the calibration process~\citep{Liu:2010,Dillon:2018}.
This approach relies on regularity in the interferometric array, leveraging the fact that for an array of identical antennas, the measured visibilities are uniquely determined (up to instrumental noise) by the corresponding baselines.
Since the physical separation between the antennas used to form a visibility uniquely determines the measured visibility (up to the antenna response), different pairs of antennas that form the same baseline will produce ``redundant'' measurements.
Arrays with elements placed on a regular grid will therefore generate several ``redundant groups'' of baselines, each of which produces multiple independent measurements of the same visibility.
These repeated measurements allow the analyst to infer the expected visibilities from the data rather than explicitly model the expected visibilities prior to calibrating the data.
In reality, the exact placement of antennas will deviate somewhat from their ideal positions in regular arrays, and there will be variations in the dish construction and feed placement from one antenna to another.
It therefore follows that there will always be some degree of nonredundancy in real interferometric arrays, and this nonredundancy may propagate to errors in the calibration solutions~\citep{Joseph:2018,Orosz:2019,Charles:2024}.
Additionally, because redundant calibration only requires that the data are consistent within each redundant group, it can at best provide a relative calibration between elements in an array.
Consequently, there are a few degeneracies in the calibration solutions that leave the absolute flux scale and phase center of the data entirely unconstrained.
The degeneracies inherent to redundant calibration solutions require that an additional ``absolute'' calibration is applied following redundant calibration~\citep{Kern:2020b}.
Absolute calibration is typically carried out in a similar way to sky-based calibration, but modifications to the gains are restricted to the degenerate subspace of the redundant calibration solutions.
Both sky-based calibration and redundant calibration are therefore subject to modeling errors, and the resulting calibration errors may be limiting systematics in the search for the cosmological 21\,cm signal~\citep{Barry:2016,Li:2018}.

In light of the increasingly urgent need for improved calibration techniques, there have been several recent efforts to improve on sky-based and redundant calibration~\citep[e.g.,][]{Byrne:2021,Ewall-Wice:2022b,Sims:2022a,Sims:2022b}.
Among the recent innovations in calibration, the unified calibration approach from~\citet{Byrne:2021} bears the closest resemblance to CorrCal.
Their approach essentially utilizes a redundant calibration likelihood that is supplemented by a sky-based prior to jointly fit for the per-antenna gains and the model visibilities in a Bayesian framework.
In other words, they enforce that the data are internally consistent while providing extra information about what they expect each unique baseline to measure.
\citet{Byrne:2021} demonstrate that their unified calibration scheme produces calibration solutions that are more accurate than those obtained through redundant calibration or sky-based calibration.
An important omission from their work, however, is the scalability of their approach and therefore whether it is a \emph{practical} tool for calibrating current and next-generation interferometric data given the immense data rates from these instruments and the computational resources available for calibration.

CorrCal is a new calibration technique that simultaneously leverages the strengths of sky-based calibration and redundant calibration in a computationally efficient framework that is resilient to modeling errors.
CorrCal achieves this through using information about the radio sky and a generalized notion of array redundancy to model the covariance between measured visibilities, optimally matching the model covariance to the sample covariance of the data.
CorrCal treats point sources and diffuse radio emission as statistically independent contributions to the measured visibilities, and in this paper we show how this delineation between sky components effectively places CorrCal as a hybrid between sky-based calibration and redundant calibration.
At a high level, the modeled point sources primarily drive the phase calibration, while the statistical model of diffuse emission provides a generalized constraint on array redundancy.
Importantly, CorrCal allows for \emph{partial} redundancy to be taken into account when calibrating the data, extending the applicability of the algorithm beyond gridded arrays.
Through modeling the expected partial redundancies, CorrCal may also see effective use on irregular arrays with dense cores, since a dense core of antennas will generate many partial redundancies between baselines.

In this paper, we extend the work of~\citet{Sievers:2017}, which initially presented CorrCal.
We provide a thorough derivation of the formalism underpinning CorrCal and detail how CorrCal achieves an efficient covariance-based approach to calibrating radio interferometric data.
We improve on the original algorithm by properly normalizing the likelihood used to obtain the calibration solutions with very little additional computational overhead.
This is an important advancement, since the calibration solutions from the previous implementation (which included only a simple regularization on the average gain phase) featured a curvature across the array due to not properly accounting for the number of redundant groups each antenna participated in.
We build on the tests that were performed in~\citet{Sievers:2017} by more systematically investigating how CorrCal responds to various modeling errors.
These are end-to-end tests of CorrCal that simulate visibilities from a model of the sky and the instrument response, apply per-antenna gains and thermal noise, then use CorrCal to infer the gains that were applied to the simulated data.
In our tests, we explore a broad range of ways in which our calibration model may be inaccurate to develop a thorough understanding of how CorrCal will perform in a variety of realistic scenarios.
All of the results presented in this paper were obtained using a new implementation of CorrCal that is publicly available on GitHub\footnote{\url{https://github.com/r-pascua/corrcal}} and contains extensive documentation, unit testing, and tutorials.

The remainder of the paper is organized in the following way.
In~\autoref{sec:FORMALISM}, we provide a high-level review of the formal foundation underlying CorrCal, highlighting the main results from the detailed calculations provided in the appendix.
In~\autoref{sec:IMPLEMENTATION}, we provide an overview of how CorrCal leverages the sparsity of the baseline--baseline covariance to attain an efficient covariance-based calibration, which is supplemented by a detailed description of the algorithm in~\autoref{sec:ALGORITHM}.
In~\autoref{sec:VALIDATION}, we present the results of a suite of validation tests designed to determine whether modeling errors manifest as calibration errors.
In~\autoref{sec:CONCLUSION}, we briefly summarize the results presented in this paper and discuss possible directions for future work.

\section{Correlation Calibration}
\label{sec:FORMALISM}
Calibration is fundamentally a problem of parameter estimation.
We have a prior belief that the data may be described by a parametric model, which we use to construct a likelihood that quantifies how well the model fits the data for a given set of model parameters.
The parameters in our model provide us with a means for translating the raw data into the physically meaningful quantities measured by the telescope.
Maximizing the likelihood with respect to the calibration parameters therefore provides us with the best estimate of the physically meaningful data products, insofar as our model is an accurate description of the data.

Traditional approaches to calibration assume that differences between the data and the model are driven by Gaussian distributed thermal noise.
Accordingly, the best-fit model can be obtained by minimizing a $\chi^2$ statistic,
\begin{equation}
    \chi^2 = \sum_k \frac{|d_k - m_k(\boldsymbol{\theta})|^2}{\sigma_k^2},
\end{equation}
where $d_k$ are the measured data (i.e., the interferometric visibilities), $m_k(\boldsymbol{\theta})$ is the model of the data characterized by parameters $\boldsymbol{\theta}$, $\sigma_k^2$ is the variance in each measurement due to thermal noise, and $k$ indexes over different data used in the fit.
Said differently, traditional approaches to calibration rely on modeling the \emph{expectation value} of the data and tweaking the model parameters to minimize the residuals in the fit given the noise level in the data.
CorrCal, on the other hand, uses a model of the expected \emph{covariance} in the data and tweaks the model parameters to obtain the best-fit to the measured covariance.

Contrary to traditional approaches to calibration, where the covariance is typically immutable, CorrCal treats the covariance as a mutable object that is varied in the calibration algorithm.
An absolutely crucial aspect of the CorrCal formalism is that we treat the sky itself as a source of randomness in the measured visibilities.
A key component of the CorrCal formalism is therefore the calculation of how spatial fluctuations in the brightness of the radio sky induce correlations between the visibilities measured by different baselines and how those correlations are modulated by the per-antenna gains.
Crucially, however, we do not compute the expected visibility--visibility covariance through empirical methods.
Instead, we use catalogued information of the radio sky to construct a statistical sky model that we propagate to a visibility--visibility covariance model according to the physics of radio interferometry.
This formal procedure relies heavily on the notion of an \emph{ensemble average}, wherein expectation values are computed by averaging over an infinite number of realizations of the radio sky.
In our formalism, bright point sources are treated as having fixed positions across all sky realizations but fluxes that fluctuate with high variance, while the remainder of the radio sky fluctuates according to a power spectrum.

\subsection{Covariance-based Optimization}
In the context of CorrCal, the covariance structure of the data is determined by the instrument response and the radio emission on the sky.
We must therefore carefully consider how these terms manifest in the covariance between measured visibilities.
Formally, we treat the data as a vector of mean-zero, correlated Gaussian random variables and maximize the likelihood that the measured data are drawn from the distribution characterized by our model covariance.
We model the data as a mixture of three independent Gaussian random components: one term characterizes thermal noise in the visibilities; one term characterizes covariances that arise from the presence of bright, unresolved emission at known locations on the sky (i.e., point sources); and the last characterizes the visibility--visibility covariance associated with diffuse emission in the form of a correlated Gaussian random field.
We emphasize that in our model, the sky itself is a Gaussian process that induces randomness in the measured visibilities, in addition to the usual radiometer noise.
Furthermore, CorrCal takes the unusual approach of treating the sky as mean zero, so our prior effectively allows for regions of negative flux on the sky.
While this modeling choice is counterintuitive and our prior technically supports unphysical radio fluxes, we show in~\autoref{sec:VALIDATION} that CorrCal is nonetheless able to obtain high quality calibration solutions in end-to-end simulations containing a realistic radio sky that includes the Galaxy and known radio point sources.
Accordingly, we defer the investigation of alternate calibration models (e.g., those that break the mean-zero assumption) to future work.

Because CorrCal takes a radically different approach to modeling the data when compared to traditional calibration strategies, the assumed statistical symmetry in the data is broken in the context of CorrCal.
The symmetry breaking is most apparent in the case of a single point source at the array's phase center, which results in purely real-valued visibilities.
In the context of CorrCal, this case is modeled as data whose variance is contained entirely within its real component.
Because we are modeling the source fluxes as random variables, with each new measurement the imaginary part of the data will remain zero while the real part will fluctuate.
More broadly speaking, a point source will typically generate unequal variance in the real and imaginary parts of the data and a nonzero covariance between the real and imaginary parts, which motivates the need to treat the real and imaginary parts of the data as separate random variables with some nontrivial covariance.
Formally, the presence of point sources breaks circular symmetry in the statistics of the data, which requires one to treat either the real and imaginary parts of the data as separate random variables or the visibility and its complex conjugate as separate random variables.
In summary, because of how we choose to model the data, a single complex-valued covariance matrix provides insufficient information for completely characterizing the statistical distribution of the data.

Of the two options presented in the previous paragraph, we opt to treat the real and imaginary parts of the data as separate random variables.
We accordingly arrange the data into a purely real-valued vector that alternates between the real and imaginary components, so that
\begin{equation}
    \mathbfit{d}^T = \Bigl(d_1^R, d_1^I, d_2^R, d_2^I, \cdots, d_N^R, d_N^I \Bigr),
\end{equation}
where $d_k^R$ indicates the real part of the $k$-th visibility, $d_k^I$ indicates the imaginary part, and $N$ is the number of measurements used for calibration.
In general, $k$ may index over any combination of baseline, time, frequency, and polarization, but for this paper we restrict our attention to baseline--baseline covariances at a single time, frequency, and polarization.
It should therefore be understood that $k$ indexes over baselines throughout the remainder of this paper.
Since we are treating the data as mean-zero, correlated Gaussian random variables, we use a multivariate Gaussian probability density as our likelihood,
\begin{equation}
    \label{eq:corrcal-likelihood}
    \mathcal{L} \propto \bigl({\rm det}\mathbf{C}\bigr)^{-1/2} \exp\biggl(-\frac{1}{2} \mathbfit{d}^T \mathbf{C}^{-1} \mathbfit{d}\biggr),
\end{equation}
where $\mathbf{C}$ is our model of the baseline--baseline covariance, which we describe in the following section.
The model covariance contains our calibration parameters, so we obtain a calibration solution by minimizing the negative log-likelihood
\begin{equation}
    \label{eq:negative-log-likelihood}
    -\log\mathcal{L} = \log{\rm det}\mathbf{C} + \mathbfit{d}^T \mathbf{C}^{-1} \mathbfit{d},
\end{equation}
where we have discarded any constant terms and overall scalings since these do not affect the location of the minimum.
The log-determinant term provides a normalization on the calibration parameters which, when taken together with the generalized $\chi^2$ term $\mathbfit{d}^T \mathbf{C}^{-1} \mathbfit{d}$, ensures that the best-fit solution produces the model covariance that is the closest match to the data covariance.

To more clearly see why our choice of likelihood ensures that CorrCal optimally matches the model covariance $\mathbf{C}$ to the data covariance, consider the derivative of the negative log-likelihood with respect to the covariance matrix elements,
\begin{equation}
    -\frac{\partial \log\mathcal{L}}{\partial C_{ij}} = \frac{1}{{\rm det}\mathbf{C}} \frac{\partial{\rm det}\mathbf{C}}{\partial C_{ij}} + \mathbfit{d}^T \frac{\partial \mathbf{C}^{-1}}{\partial C_{ij}} \mathbfit{d}.
\end{equation}
The derivative of the determinant is given by Jacobi's formula, so that
\begin{equation}
    \frac{\partial {\rm det} \mathbf{C}}{\partial C_{ij}} = {\rm det}\mathbf{C} \bigl(\mathbf{C}^{-1}\bigr)_{ij}.
\end{equation}
Since the derivative of the inverse can be rewritten via
\begin{equation}
    \frac{\partial \mathbf{C}^{-1}}{\partial C_{ij}} = -\mathbf{C}^{-1} \frac{\partial \mathbf{C}}{\mathbf{C}_{ij}} \mathbf{C}^{-1},
\end{equation}
we can rewrite the derivative of the generalized $\chi^2$ term as
\begin{equation}
    \mathbfit{d}^T \frac{\partial \mathbf{C}^{-1}}{\partial C_{ij}} \mathbfit{d} = -\sum_{m,n} d_m \bigl(\mathbf{C}^{-1}\bigr)_{mi} \bigl(\mathbf{C}^{-1}\bigr)_{jn} d_n.
\end{equation}
This result, taken together with Jacobi's formula, allows us to express the derivative of the negative log-likelihood with respect to the covariance as
\begin{equation}
    -\frac{\partial\log\mathcal{L}}{\partial\mathbf{C}} = \mathbf{C}^{-1} - \mathbf{C}^{-1} \mathbfit{dd}^T \mathbf{C}^{-1},
\end{equation}
and therefore the negative log-likelihood is minimized when $\mathbf{C} = \mathbfit{dd}^T$.
The expected data covariance (over all realizations of the radio sky) is $\bigl\langle \mathbfit{dd}^T\bigr\rangle$, where $\langle \cdots \rangle$ indicates an ensemble average, so our choice of likelihood produces calibration solutions that, on average, optimally match our model covariance to the data covariance.
While $\mathbf{C} = \mathbfit{dd}^T$ is not a formally admissible solution, because this best-fit covariance is not invertible (since $\mathbf{C} = \mathbfit{dd}^T$ is rank 1), the result carries a useful conceptual interpretation.
By choosing to maximize the likelihood in~\autoref{eq:corrcal-likelihood}, we are therefore mathematically encoding our decision to perform calibration by fitting a model of correlations between visibilities to the observed correlations in the data.
Furthermore, the normalization term is crucial for ensuring that the likelihood is maximized when the model covariance matches the data covariance---a different choice of normalization would result in a different optimal covariance.

The calculation provided in the previous paragraph suggests that the optimal solution for our covariance-based approach is to make the covariance model exactly match the observed data covariance.
In the context of calibrating radio interferometric data, however, we are instead solving a \emph{constrained} optimization problem: we are combining covariance optimization with the assumptions of radio interferometry.
When we perform calibration we are not fitting the model covariance matrix elements directly, but rather each matrix element is a function of the calibration parameters.
While this is partially due to the fact that fitting the individual matrix elements is an extremely under-constrained problem, our main motivation is that the covariance matrix elements on their own do not provide us with a link between the raw data and the physically meaningful telescope outputs.
In the next part of this section, we provide an overview of how the model covariance takes on a sparse form that makes manifest CorrCal's status as a hybrid between sky-based calibration and redundant calibration.
For completeness, we provide a detailed derivation of the sparse covariance representation in~\autoref{sec:SOURCE-COVARIANCE} and~\autoref{sec:DIFFUSE-COVARIANCE}.

\subsection{Correlating Visibilities}
\label{sec:MODEL-COVARIANCE}
We begin the calculation of the model covariance $\mathbf{C}$ with the model $m_k$ for the visibilities,
\begin{equation}
    \label{eq:model-visibility}
    m_k = G_k \int A_k(\bvec{r}) I(\bvec{r}) e^{-i2\pi\nu \mathbfit{b}_k \cdot \bvec{r}/c} d\Omega + n_k,
\end{equation}
where $A_k(\bvec{r})$ is the primary beam for baseline $k$ in direction $\bvec{r}$, $I(\bvec{r})$ is the intensity of radio emission on the sky, $\mathbfit{b}_k = \mathbfit{x}_{k_2} - \mathbfit{x}_{k_1}$ (where $\mathbfit{x}_i$ is the position of antenna $i$) is the spatial separation between antennas $k_1$ and $k_2$, $\nu$ is the observed frequency, $c$ is the speed of light in vacuum, $d\Omega$ is the differential solid angle element, $G_k \equiv g_{k_1} g_{k_2}^*$ is the $k$-th entry of the ``gain matrix'', $g_{k_1}$ and $g_{k_2}$ are the direction-independent gains, and $n_k$ is radiometer noise.
Because we need to treat the real and imaginary parts of the data as separate random variables, we arrange the model visibilities into a purely real-valued vector $\mathbfit{m}$ so that
\begin{equation}
    \mathbfit{m}^T = \Bigl(m_1^R, m_1^I, \cdots, m_N^R, m_N^I\Bigr),
\end{equation}
and compute the model covariance $\mathbf{C}$ via $\mathbf{C} = \bigl\langle \mathbfit{mm}^T \bigr\rangle$, where we again stress that the ensemble average is taken over all realizations of radiometer noise and all radio sky realizations.
Because we arrange the model visibilities into alternating real and imaginary parts, the model covariance consists of $2 \times 2$ blocks,
\begin{equation}
     \mathbf{C}  = 
    \begin{pmatrix}
          \mathbf{C}_{11} & \cdots & \mathbf{C}_{1N} \\
          \vdots & \ddots & \vdots \\
          \mathbf{C}_{N1} & \cdots & \mathbf{C}_{NN}
    \end{pmatrix},
\end{equation}
with each $2 \times 2$ block $\mathbf{C}_{ij}$ taking the form
\begin{equation}
    \label{eq:covariance-block}
    \mathbf{C}_{ij} =
    \begin{pmatrix}
        \bigl\langle m_i^R m_j^R \bigr\rangle & \bigl\langle m_i^R m_j^I \bigr\rangle \\
        \bigl\langle m_i^I m_j^R \bigr\rangle & \bigl\langle m_i^I m_j^I \bigr\rangle
    \end{pmatrix}.
\end{equation}
Considering both the above equation and~\autoref{eq:model-visibility}, constructing our model covariance requires computing the sky covariance and propagating the result through the measurement equation, which we outline below and derive in detail in~\autoref{sec:SOURCE-COVARIANCE} and~\autoref{sec:DIFFUSE-COVARIANCE}.

We model the sky intensity $I(\bvec{r})$ as a mean-zero Gaussian random field, which we decompose into a diffuse component and a point source component via
\begin{equation}
    \label{eq:sky-components}
    I(\bvec{r}) = \mathcal{D}(\bvec{r}) + \sum_j s_j \delta^D(\bvec{r} - \bvec{r}_j),
\end{equation}
where the Dirac delta $\delta^D(\cdots)$ encodes the assumption that we know precisely where each source is located.
We assume that the diffuse component $\mathcal{D}(\bvec{r})$ is uncorrelated with each source, and that the flux density $s_j$ of a source is uncorrelated with that of any other source.
In other words, we treat the sky as a mixture of mutually independent Gaussian random components.
The sky covariance is therefore
\begin{align}
    \nonumber
    \bigl\langle I(\bvec{r}) I(\bvec{r}') \bigr\rangle &= \bigl\langle \mathcal{D}(\bvec{r}) \mathcal{D}(\bvec{r}') \bigr\rangle \\
    &+ \sum_j S_j^2 \delta^D(\bvec{r} - \bvec{r}_j)\delta^D(\bvec{r}' - \bvec{r}_j),
    \label{eq:sky-cov}
\end{align}
where $S_j^2$ is the variance in the flux density for source $j$.
Because we assume that radiometer noise is uncorrelated with the radio sky,~\autoref{eq:sky-cov} immediately tells us that the visibility--visibility covariance separates into three components,
\begin{equation}
    \label{eq:covariance-components}
    \mathbf{C} = \mathbf{N} + \mathbf{GDG}^T + \mathbf{GSG}^T,
\end{equation}
where $\mathbf{N}$ captures the per-baseline variance due to radiometer noise, $\mathbf{D}$ captures the covariance induced by the diffuse sky signal, $\mathbf{S}$ captures the covariance induced by point sources, and $\mathbf{G}$ is a block-diagonal matrix containing the gains.
As we show in detail in~\autoref{sec:SOURCE-COVARIANCE} and~\autoref{sec:DIFFUSE-COVARIANCE}, it turns out that the covariance is sparse in the sense that one does not need the full $2N_{\rm bl} \times 2N_{\rm bl}$ covariance matrix elements to completely describe the covariance---all of the information is captured in far fewer numbers.
In the remainder of this section, we provide physical arguments to motivate this key result from~\autoref{sec:SOURCE-COVARIANCE} and~\autoref{sec:DIFFUSE-COVARIANCE}.

The motivation for the sparsity of the thermal noise covariance $\mathbf{N}$ is very straightforward.
Radiometer noise is independent between baselines, and in a given visibility the noise in the real component is independent from that in the imaginary component.
The thermal noise covariance is therefore manifestly diagonal---a thermal noise fluctuation in one visibility is completely independent of that in another visibility.
As a result, we only need to keep track of the $N_{\rm bl}$ entries along the diagonal of the thermal noise covariance.

For the source covariance, the key idea is that each source independently contributes to the total covariance (which is why there is a single sum over sources in~\autoref{eq:sky-cov} instead of a double sum) and the measurement equation tells us the visibility phase associated with each source as well as how much the source brightness is attenuated by the beam.
Because each term in~\autoref{eq:covariance-block} is separable in baseline indices, and because the sky covariance involves a single sum over sources, the source covariance may be written as a matrix product via $\mathbf{S} = \mathbf{\Sigma\Sigma}^T$.
Each column of the source matrix $\mathbf{\Sigma}$ encodes the full array response to a particular point source, so the source covariance is completely characterized by $N_{\rm bl} \times N_{\rm source}$ numbers, where $N_{\rm source}$ is the number of sources included in the source covariance model.

For the diffuse covariance, the key idea is that the majority of the information in the two-point correlation function $\bigl\langle\mathcal{D}(\bvec{r})\mathcal{D}(\bvec{r}')\bigr\rangle$ is captured in the isotropic power spectrum, which characterizes how the sky covariance changes as a function of angular scale.
One of the core principles of radio interferometry is that a visibility is most sensitive to structures on the sky with an angular frequency determined by the baseline, $\mathbfit{u} = \mathbfit{b} / \lambda$, where $\lambda$ is the observed wavelength.
Because the covariance of an isotropic signal at one angular scale is independent of that at another angular scale, whether the diffuse signal induces a correlation between a pair of visibilities depends on whether those visibilities are sensitive to the same angular scales.
Consequently, many of the entries in the diffuse covariance will be zero, since only visibilities with sufficiently similar baselines will correlate with one another.
We therefore do not need to keep track of all $N_{\rm bl} \times N_{\rm bl}$ terms in the diffuse covariance, but rather keep track of only the relatively few pairs of baselines that do correlate with one another.

\begin{figure*}
    \includegraphics[width=\textwidth]{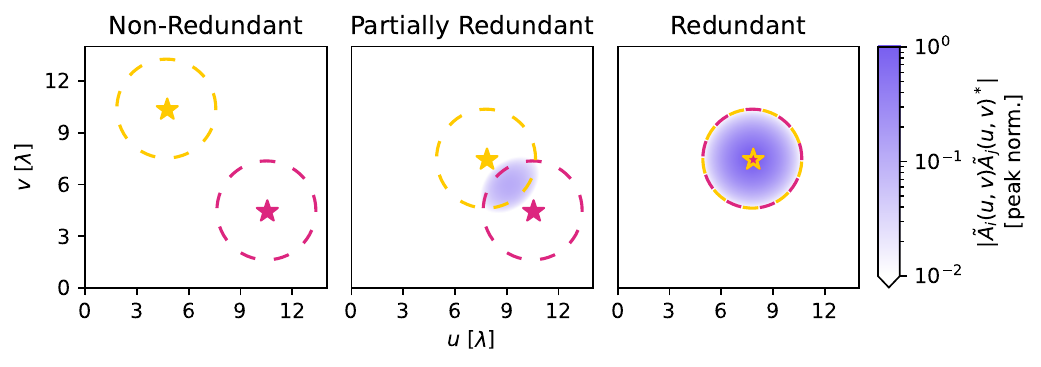}
    \caption{
        \label{fig:redundancy-example}
        Schematic showing the product of the beam kernels in the overlap integral that is used to compute the diffuse covariance matrix elements.
        The dashed lines show the expected position of the first null in the beam kernel, the stars indicate the central $uv$ modes sampled by the different baselines, and the color scale indicates the absolute value of the peak-normalized product of beam kernels at each point in the $uv$-plane.
        The left panel shows a pair of baselines that are considered non-redundant; the center panel shows a pair of baselines that are partially redundant; and the right panel shows a pair of baselines that are perfectly redundant.
    }
\end{figure*}

Because the diffuse covariance depends on baseline similarity, it is fairly plain to see that the diffuse covariance serves as the redundant calibration analog in CorrCal.
A more subtle consequence, however, is that the diffuse covariance can also capture partial redundancy between baselines.
The key to understanding how we can leverage partial redundancy relies on the fact that an interferometer does not measure a single angular scale on the sky---the beam smears out a visibility's sensitivity to modes in the $uv$-plane to some finite region centered on $\mathbfit{u} = \mathbfit{b} / \lambda$.
We refer to this beam-induced smearing as the \emph{beam kernel}, $\tilde{A}_k(\mathbfit{u})$, and we show in~\autoref{sec:DIFFUSE-COVARIANCE} that the beam kernel is an essential ingredient in computing the diffuse covariance.
More precisely, how sensitive a baseline is to a particular $uv$-mode is quantified by the beam kernel shifted by an amount determined by the baseline, $\tilde{A}_k(\mathbfit{u} - \mathbfit{b}_k/\lambda)$, and the whether two baselines will correlate depends on whether their shifted beam kernels overlap.
We show this schematically in~\autoref{fig:redundancy-example} by plotting the product of shifted beam kernels for a pair of nonredundant baselines, a pair of partially redundant baselines, and a pair of perfectly redundant baselines.
In practical terms, when the baselines are sorted by redundancy, the diffuse covariance takes on a block structure, with the highest amplitude blocks on the diagonal and only a handful of nonzero, lower amplitude blocks in the off-diagonal.

An important use-case for CorrCal is the application to gridded interferometric arrays, such as HERA, CHORD, and HIRAX.
For these arrays, correlations within a redundant group greatly outweigh correlations between baselines from different redundant groups, so we may reasonably approximate the diffuse covariance by ignoring the partial redundancy terms.
In this approximation, the diffuse covariance reduces to a block-diagonal matrix (under a redundancy-based baseline ordering), and each block may be independently factored via eigendecomposition.
Accordingly, in this limit we may write $\mathbf{D} = \mathbf{\Delta\Delta}^T$, where the \emph{diffuse matrix} $\mathbf{\Delta}$ contains only $N_{\rm bl} \times N_{\rm eig}$ nonzero values, with $N_{\rm eig}$ eigenmodes used to factor the covariance within each redundant group.
This factorization of the diffuse covariance allows for the model covariance to be written in sparse form as
\begin{equation}
    \label{eq:full-sparse-covariance}
    \mathbf{C} = \mathbf{N} + \mathbf{G\Delta\Delta}^T\mathbf{G}^T + \mathbf{G\Sigma\Sigma}^T\mathbf{G}^T.
\end{equation}
Broadly speaking, the source matrix $\mathbf{\Sigma}$ serves as the sky-based calibration analog in CorrCal, while the diffuse matrix $\mathbf{\Delta}$ serves as the redundant calibration analog.
Furthermore,~\autoref{eq:full-sparse-covariance} shows that the model covariance (at least in this limiting case) can be characterized in $N_{\rm bl}(1 + N_{\rm source} + N_{\rm eig})$ terms rather than the full $N_{\rm bl}^2$ terms in the dense covariance.
Finally, although~\autoref{eq:full-sparse-covariance} was obtained by disregarding partial redundancies in the diffuse covariance, the more general case should also permit a similar factorization of the diffuse covariance where $\mathbf{\Delta}$ is block-triangular instead of block-diagonal.

\section{Implementation}
\label{sec:IMPLEMENTATION}
Thus far we have established that the guiding principle behind CorrCal is to fit a model of correlations between visibilities to the correlations observed in the data.
As a covariance-based calibration technique, it is reasonable to assume that CorrCal would be computationally inefficient for large arrays---inverting an $N \times N$ matrix scales as $\mathcal{O}(N^3)$, and for a baseline--baseline covariance $N$ scales with the square of the number of antennas in the array.
Recall, however, that by choosing to treat the sky as a Gaussian random field plus a collection of point sources, the baseline--baseline covariance matrix has a sparse representation.
In this section, we outline how the sparsity of the model covariance may be leveraged to enable a computationally efficient covariance-based calibration of interferometric arrays with a large number of antennas.
The discussion in this section is supplemented in~\autoref{sec:ALGORITHM} by a detailed description of the algorithm as implemented in the GitHub repository\footnote{\url{https://github.com/r-pascua/corrcal}}.

\subsection{Sparse Two-Level Covariance}
\label{sec:SPARSE-TWO-LEVEL}
The core functionality of CorrCal revolves around the notion of a ``sparse two-level'' covariance, which alludes to the fact that both the covariance and some of its constituent matrices are sparse.
Recall from~\autoref{eq:full-sparse-covariance} that the baseline--baseline covariance $\mathbf{C}$ is fully characterized by four matrices: $\mathbf{N}$, $\mathbf{G}$, $\mathbf{\Delta}$, and $\mathbf{\Sigma}$.
Of these four matrices, only the source matrix $\mathbf{\Sigma}$ is not sparse.
Since the noise matrix $\mathbf{N}$ is diagonal and the gain matrix $\mathbf{G}$ is block-diagonal with each block a $2 \times 2$ matrix, calculations involving these matrices are straightforward to implement efficiently.
While the sparsity of the diffuse matrix $\mathbf{\Delta}$ depends on the array configuration, much of the diffuse matrix will be zero in general, since only a small fraction of the unique baselines will have overlapping beam kernels.
Furthermore, by sorting the baseline axis according to redundancy, the diffuse covariance takes on a block structure that admits a block-triangular representation for $\mathbf{\Delta}$.
One therefore only needs to keep track of the nonzero blocks in $\mathbf{\Delta}$ and restrict matrix operations to only accumulate contributions from those nonzero blocks.
The ``sparse two-level'' structure is then implemented through routines for interacting with the constituent matrices of the model covariance and custom kernels for sparse matrix operations involving the diffuse matrix.

The factorization of the model covariance according to~\autoref{eq:full-sparse-covariance} allows for efficient calculation of the log-likelihood and its gradient, which form the crux of the calibration loop.
Central to both of these routines is the inversion of the model covariance, which is made efficient through a two-stage application of the Woodbury identity,
\begin{equation}
    \bigl(\mathbf{A} + \mathbf{UV}^T\bigr)^{-1} = \mathbf{A}^{-1} - \mathbf{A}^{-1} \mathbf{U} \bigl(\mathbf{I} + \mathbf{V}^T \mathbf{A}^{-1} \mathbf{U}\bigr)^{-1} \mathbf{V}^T \mathbf{A}^{-1},
\end{equation}
where $\mathbf{I}$ is the identity matrix.
At a high level, this process first inverts $\tilde{\mathbf{C}} = \mathbf{N} + \mathbf{G\Delta\Delta}^T\mathbf{G}^T$, then inverts $\mathbf{C} = \tilde{\mathbf{C}} + \mathbf{G\Sigma\Sigma}^T\mathbf{G}^T$.
By taking the Cholesky factorization of the small matrices $\mathbf{I} + \mathbf{V}^T\mathbf{A}^{-1} \mathbf{U}$ that appear in each stage, it is possible to write the inverse covariance in a sparse form similar to~\autoref{eq:full-sparse-covariance} via
\begin{equation}
    \label{eq:sparse-inverse-covariance}
    \mathbf{C}^{-1} = \mathbf{N}^{-1} - \bar{\mathbf{\Delta}} \bar{\mathbf{\Delta}}^T - \bar{\mathbf{\Sigma}} \bar{\mathbf{\Sigma}}^T,
\end{equation}
where $\bar{\mathbf{\Delta}}$ and $\bar{\mathbf{\Sigma}}$ have the same shapes as $\mathbf{\Delta}$ and $\mathbf{\Sigma}$, respectively.
Ultimately, this process allows the calibration loop calculations to be written solely in terms of $\mathbf{N}, \mathbf{G}, \mathbf{\Delta},$ and $\mathbf{\Sigma}$ (and the analogous matrices in~\autoref{eq:sparse-inverse-covariance}).
In addition, the log-determinant term in~\autoref{eq:negative-log-likelihood} can be efficiently accumulated in the inversion procedure by using the Matrix Determinant Lemma,
\begin{equation}
    \det\bigl(\mathbf{A} + \mathbf{UV}^T\bigr) = \det\bigl(\mathbf{I} + \mathbf{V}^T \mathbf{A}^{-1} \mathbf{U}\bigr) \det\mathbf{A}.
\end{equation}
A more detailed description of the algorithm (in the case where partial redundancies are ignored) is provided in~\autoref{sec:ALGORITHM}.

\subsection{Performance}
\label{sec:PERFORMANCE}

\begin{figure*}
    \includegraphics[width=\textwidth]{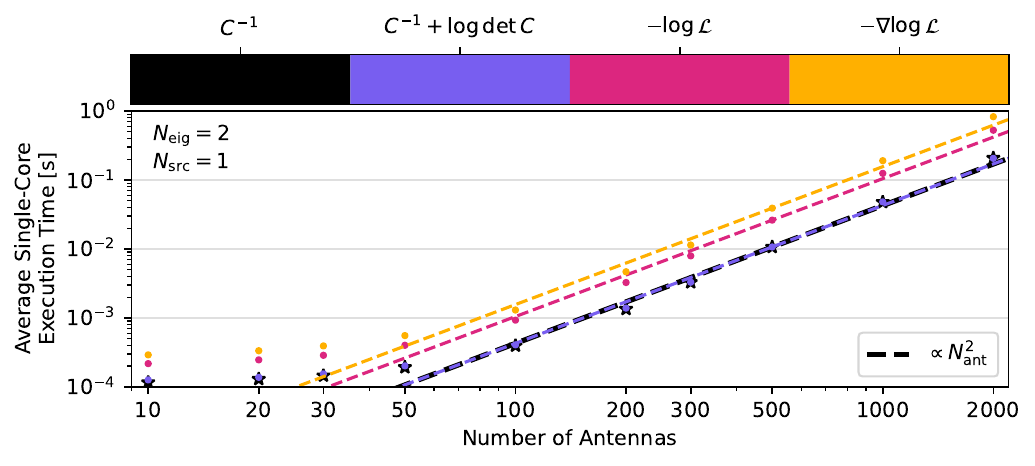}
    \caption{
        Results of a benchmark test for the single core execution time (using an Intel i7-8565U processor) of the most computationally intensive tasks in the CorrCal algorithm, shown as a function of array size: black shows the average time to invert the covariance; purple shows the time to invert the covariance and simultaneously accumulate the log-determinant; pink shows the total time for computing the negative log-likelihood; and yellow shows the total time to compute the gradient of the negative log-likelihood.
        For arrays with less than roughly 100 antennas, the computational cost is dominated by overheads.
        For larger arrays, the computational cost of all the tasks investigated here roughly scales with the square of the number of antennas, or linearly with the number of baselines.
        The dashed lines indicate execution times that scale with the square of the number of antennas, normalized to the measured execution time for a 500-element array.
        \label{fig:timing-test-results}
    }
\end{figure*}

In this section, we briefly discuss the computational complexity of the algorithm reviewed in the previous section.
In our analytic scaling calculations, we assume that the number of baselines $N_{\rm bl}$ greatly exceeds the number of sources $N_{\rm src}$ and the number of complex eigenmodes $N_{\rm eig}$ used to construct the source matrix $\mathbf{\Sigma}$ and the diffuse matrix $\mathbf{\Delta}$, respectively.
Under this assumption, the rate limiting step in our implementation of the likelihood evaluation algorithm scales as $\mathcal{O}\bigl(N_{\rm bl} (N_{\rm src}^2 + N_{\rm eig}^2)\bigr)$.
The rate limiting step in the gradient calculation also scales with the number of baselines, but with a different prefactor that is evidently larger than that of the likelihood evaluation, as seen in~\autoref{fig:timing-test-results}.

In~\autoref{fig:timing-test-results}, we show the average single-core execution time for various components of the CorrCal algorithm as a function of number of antennas.
We carried out the performance test on a laptop using a single Intel i7-8565U processor with minimal background processes active.
For each data point in~\autoref{fig:timing-test-results}, we ran 10 trials and estimated the average walltime for inverting the covariance without computing the log-determinant, inverting the covariance while also accumulating the log-determinant, computing the negative log-likelihood, and computing the gradient of the negative log-likelihood.
For the arrays with less than 100 antennas, we ran 1000 iterations per trial, and for the arrays with at least 100 antennas we reduced this to 100 iterations per trial.
The variance in the estimated compute times is negligible for the majority of test cases---the only exceptions were the tests with a small number of antennas, where overheads dominated the runtime.
We find that the compute times roughly scale with the number of baselines $N_{\rm bl} \sim N_{\rm ant}^2$ and that for an array with 1000 antennas the negative log-likelihood can be computed in about 0.125\,s and the gradient can be computed in about 0.19\,s when calibrating with a single source and single eigenmode.
This performance was obtained without any major optimizations beyond carefully choosing the order of operations.
Substantial improvements in performance may therefore be obtained by implementations that more efficiently manage data transfer or by transferring the calculations to a GPU-based implementation.

\vspace{\baselineskip}
\section{Validation}
\label{sec:VALIDATION}
\begin{figure*}
    \includegraphics[width=\textwidth]{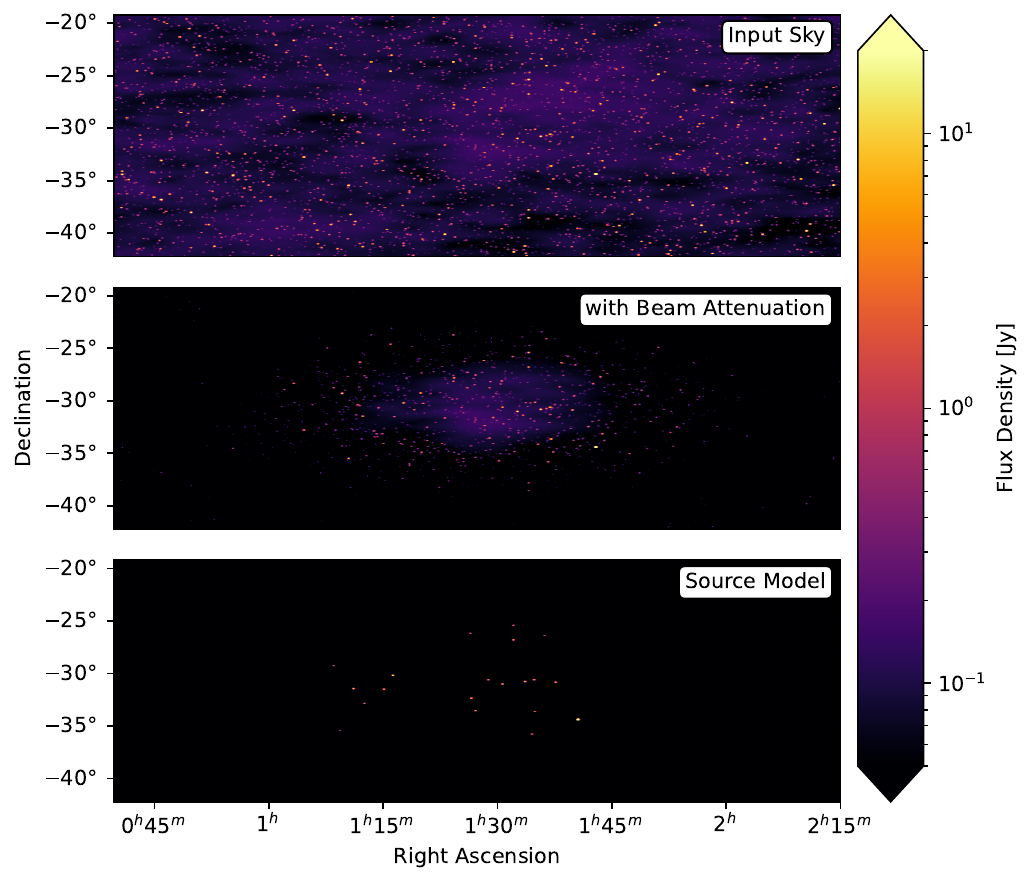}
    \caption{
        A representative view of the simulated sky, instrument response, and source model used in the tests in~\autoref{sec:VALIDATION}.
        The point sources shown in each panel have been smoothed with a $1\arcmin$ Gaussian kernel.
        The top panel shows the sky input to the visibility simulations for one of the ``bright'' fields with the diffuse component scaled down to increase the contrast with the point sources.
        The middle panel shows the same field, but with the primary beam attenuation applied.
        The bottom panel provides a representative view of the sources that would be included when modeling the source matrix $\mathbf{\Sigma}$.
    }
    \label{fig:sky-model}
\end{figure*}

In order to test CorrCal's efficacy, we perform a variety of test calibration runs with simulated data.
These tests are intended to assess how the quality of calibration solutions obtained with CorrCal are related to modeling errors.
In~\autoref{sec:PTSRC-TESTS}, we discuss the results of tests investigating how point source modeling errors affect the quality of calibration solutions.
In~\autoref{sec:NONRED-TESTS}, we investigate how CorrCal performs when applied to data with known positional nonredundancy (i.e., when the antenna positions deviate slightly from a regular grid placement).
In~\autoref{sec:REALISTIC-SKY}, we apply CorrCal to simulated data featuring a realistic sky model to assess whether non-Gaussianities in the diffuse sky signal produce significant calibration errors.
For all of the simulations, we use an Airy disk for the antenna response (i.e., a diffraction-limited beam for a uniformly illuminated dish), since it strikes a balance between moderate realism and computational simplicity.
We perform all of our simulations at an observed frequency of 150\,MHz, since this is contained in the HERA observing band and minimizes the computational resources required to simulate visibilities for the diffuse sky component.\footnote{At high frequencies, the short fringe spacings require extremely high resolution pixellizations of the sky to avoid aliasing issues.}
In~\autoref{fig:sky-model}, we provide a view of a typical simulated sky used for the tests described in~\autoref{sec:PTSRC-TESTS} and~\autoref{sec:NONRED-TESTS}.
All of the visibility simulations were performed with the \texttt{matvis} package~\citep{Kittiwisit:2025} using the wrappers available in \texttt{hera\_sim}.\footnote{\url{https://github.com/HERA-Team/hera_sim}}

\subsection{Calibrating with Incomplete Information}
\label{sec:INCOMPLETE-MODEL}
Practical applications of CorrCal will include only a small number of point sources in the source matrix, with the remainder added to the power spectrum used to compute the diffuse matrix.
This is partly due to computational concerns, since the rate limiting step in the likelihood evaluation scales with the square of the number of modeled sources.
The main motivation, however, is that there are diminishing returns to including more sources in the source matrix---including more sources beyond the brightest few offers little improvement in the calibration quality.
It is therefore pragmatic to limit the number of sources in the source matrix, as including more than the brightest few sources will significantly increase the computational burden while only marginally improving the calibration quality.
This modeling decision has a subtle effect on the recovered calibration solutions, and in the remainder of this section we discuss these effects and describe how they are accounted for when analyzing the simulation results.

Using an incomplete source catalog to construct the source matrix primarily has the effect of introducing a small spatial phase gradient in the calibration solutions.
We defer a more detailed discussion of this effect to~\autoref{sec:PHASE-SLOPE} and instead provide a heuristic argument here along with a summary of our findings from~\autoref{sec:PHASE-SLOPE}.
Broadly speaking, by choosing to only include a select few point sources in the source matrix, we are asserting that the phase structure in the visibilities is determined primarily by those point sources.
If the remaining sources do, however, contribute meaningfully to the phase structure in the data, then the calibration algorithm will try to compensate for the difference by moving around the flux on the sky.
This results in a spatial phase gradient in the best-fit gains, since a spatial phase gradient in the gains is equivalent to an overall shift in the positions of sources.
How large the resulting phase gradient is depends on how much of the total observed flux is accounted for in the source matrix---if the source model contains the majority of the observed flux from point sources, then the resulting phase gradient will be small.
This stance is corroborated by the results discussed in~\autoref{sec:PHASE-SLOPE}, which show that the phase gradients measured in simulation correspond to pointing errors smaller than the synthesized beam for the array used in the simulations.

There is an additional subtlety in the calibration solutions that is linked to the use of a statistical calibration model.
The absolute flux scale of the calibrated visibilities is determined by the amount of power in the model covariance, with some of that power coming from terms in the source matrix, some coming from the power spectrum used to construct the diffuse matrix, and the remainder coming from the provided estimate of the thermal noise variance.
Of particular importance here is the power spectrum term, as it informs us of how much power we expect to measure in each baseline on average.
For a single realization of the sky, however, there may be regions of the sky where fluctuations at a particular angular scale are enhanced or suppressed relative to the expectation from the modeled power spectrum.
As a result, while we expect our calibration solutions to place the visibilities on the correct flux scale on average, there may be some deviation from the correct value depending on the patch of sky overhead.
Concretely, the best-fit gain amplitudes may feature an overall bias that depends on how the total power in the covariance model compares to the total observed power.
In all of our tests, we have observed this bias to be the same from one antenna to another (but with variable amplitude across sky realizations).
We therefore conclude that any bias due to this effect can be relatively easily removed by ensuring the autocorrelation amplitudes match the expected amplitudes.

In summary, due to incomplete accounting of point sources and sample variance effects, we expect the calibration solutions to have a small (sub-synthesized beam equivalent) phase gradient and an easily correctable amplitude bias.
Importantly, these features are \emph{not} degeneracies in the calibration solutions---they have preferred values that are some complicated function of the calibration model and the observed sky.
While we expect the phase gradients to be small and that the amplitude biases may be easily corrected, the presence of these features can complicate the interpretation of our simulation results if unaccounted for when analyzing the results.
To that end, in the tests that follow, we leverage our knowledge of the true gains $g_a^{\rm true}$ to remove any extant phase gradient and amplitude bias through a simple absolute calibration step.
Specifically, we transform the best-fit gains $\hat{g}_a$ via
\begin{equation}
    \label{eq:gain-correction}
    \hat{g}_a' = A \hat{g}_a \exp\bigl(i\nabla\boldsymbol{\Phi} \cdot \mathbfit{x}_a\bigr),
\end{equation}
where $\mathbfit{x}_a$ is the position of antenna $a$ in a local coordinates.
The amplitude correction $A$ is computed by computing the ratio of the average true gain amplitude against the average best-fit gain amplitude,
\begin{equation}
    \label{eq:amplitude-correction}
    A = \frac{\sum_a \big|g_a^{\rm true}\big|}{\sum_a \big|\hat{g}_a\big|},
\end{equation}
which is analogous to comparing the average expected autocorrelation against the average calibrated autocorrelation.
The phase gradient $\nabla\boldsymbol{\Phi}$ is obtained from a linear least-squares fit to the phase errors in the calibrated visibilities, $\Delta\phi_k \equiv {\rm arg}\bigl(\hat{G}_k^* G_k^{\rm true}\bigr)$, by solving the system of equations
\begin{equation}
    \label{eq:phase-correction}
    \Delta\phi_k = \mathbfit{b}_k \cdot \nabla\boldsymbol{\Phi}.
\end{equation}
In the following sections, our analysis focuses on the corrected gains $\hat{g}_a'$ so that we can accurately measure the scatter in the calibration solutions.

\subsection{Sky Modeling Errors}
\label{sec:PTSRC-TESTS}
For this set of tests, we ran an ensemble of simulations where each simulation features a different level of inaccuracy in modeling the array response to point sources.
We parametrize the source model accuracy by the number of sources included in the calibration model and the average error on the model source flux densities.
For a test with $N_{\rm cal}$ calibration sources and an average fractional flux error $0 < \sigma < 1$, we construct the source matrix by taking the $N_{\rm cal}$ sources with the brightest observed flux and computing the source matrix elements according to~\autoref{eq:source-matrix} with $S_j^{\rm model} = S_j^{\rm true}(1+\varepsilon)$, where $\varepsilon \sim \mathcal{N}(0,\sigma^2)$.
The complete set of tests explores the outer product of the parameter options listed in~\autoref{table:src-model-params}.
Generically, a greater number of sources and a lower average error on the source flux densities corresponds to a more accurate source model.
For all of the tests in this section, we use a 19-element hexagonal array of antennas with a nearest-neighbor separation of 14.6\,m and a dish diameter of 14\,m.

\begin{deluxetable}
    {cc}
    \tablehead{\colhead{Test Hyperparameter} & \colhead{Options}}
    \tablewidth{\columnwidth}
    \tablecaption{
        \label{table:src-model-params} Source Model Accuracy Test Parameters
    }
        \startdata
            Average Flux Error [\%] & (0, 1, 2, 5, 10, 20) \\
            Number of Calibration Sources & (1, 5, 10, 20) \\
        \enddata
\end{deluxetable}

Because we will never calibrate with a complete sky model, as discussed in the previous section, we investigate how the quality of the calibration depends on the availability of bright point sources on the sky.
To that end, we experiment with 10 realizations each of three different distributions of point sources on the sky.
Each of these distributions uses the distribution of source fluxes reported in~\citet{Franzen:2019}, but enforces different cutoffs in the maximum source flux at 150\,MHz: the ``quiet'' field only includes sources up to 1\,Jy; the ``average'' field includes sources up to 100\,Jy; and the ``bright'' field includes sources up to 100\,Jy with an additional bright source whose flux ranges between 1\,kJy and 10\,kJy.
For each of these fields, we only simulate sources in a $30^\circ \times 30^\circ$ patch of sky centered on zenith and enforce a common minimum flux cutoff of 100\,mJy.
This yields several thousand sources per simulation and should not produce qualitatively different results than simulating sources over the full sky since the simulated antenna response $30^\circ$ away from zenith is suppressed by roughly a factor of $10^3$.

\begin{deluxetable}{cc}
    \tablehead{\colhead{Parameter} & \colhead{Distribution}}
    \tablewidth{\columnwidth}
    \tablecaption{
        \label{table:sim-gain-parameters} Simulated Gain Parameters
    }
    \startdata
        True Gain Amplitude, $g^{\rm true}_a$ & $\mathcal{N}(1, 0.1^2)$ \\
        True Gain Phase, $\phi^{\rm true}_a$ & Uniform$(0,2\pi)$ \\
        Initial Gain Amplitude Error, $\frac{g^{\rm init}_a}{g^{\rm true}_a}$ & $\mathcal{N}(1,0.05^2)$ \\
        Initial Gain Phase Error, $\phi^{\rm init}_a - \phi^{\rm true}_a$ & $\mathcal{N}(0,0.02^2)$
    \enddata
\end{deluxetable}

Each simulation also contains a diffuse sky component that is a realization of a Gaussian random field with an angular power spectrum $C_\ell \propto (1 + \ell)^{-2}$.
To ensure that the diffuse sky component is non-negative, we add a monopole $T_0$ to the randomly generated diffuse map.
We additionally rescale the diffuse component by a factor $\mathcal{A}$ so that the maximum brightness in the generated map is 500\,K, which is roughly comparable to the sky brightness away from the plane of the galaxy at 150\,MHz.
Accounting for the rescaling and monopole, the power spectrum of the simulated diffuse sky is
\begin{equation}
    C_\ell^{\rm sim} = \mathcal{A}^2 C_\ell^{\rm input} \bigl(1 + 4\pi T_0^2 \delta_{\ell 0}\bigr).
\end{equation}
We use this diffuse power spectrum, after converting it from temperature units to telescope units, when computing the diffuse matrix $\mathbf{\Delta}$.

The set of tests may be thought of as being split up into different ``trials,'' where each trial consists of 100 different calibration runs with different realizations of thermal noise.
For each trial, we generate one set of per-antenna gains and one set of initial guesses for the gains that are fairly close to the true gains.
The true gain phases are uniformly distributed on $[0,2\pi)$, and the true gain amplitudes are normally distributed about unity with a standard deviation of 0.1.
The phases of the initial guesses are normally distributed about the true phases with a standard deviation of 0.02 radians, and the ratios between the true amplitudes and the amplitudes of the initial guesses are normally distributed about unity with a standard deviation of 0.05.
These parameters are summarized in~\autoref{table:sim-gain-parameters}.
Since we would like to minimize the effect of thermal noise on the quality of the calibration solutions, we simulate radiometer noise such that the signal-to-noise ratio in the visibilities is roughly several hundred on average.
Additionally, to minimize the errors from sample variance, we perform 10 trials for each choice of source modeling error and sky realization, providing us with a combined 720,000 calibration solution samples.

To assess the quality of the calibration solutions, we primarily rely on a chi-squared per degree of freedom statistic $\chi^2/{\rm DoF}$, as well as a $p$-value statistic, both of which are computed once per calibration run.
We choose to use these statistics as our metric for calibration quality, rather than the distributions of the calibration errors themselves, because $\chi^2$ is normalized against the noise amplitude, which allows for straightforward comparison between trials.
We compute $\chi^2$ via
\begin{equation}
    \label{eq:src-test-chisq}
    \chi^2 = \sum_k \frac{\big|V_k^{\rm cal} - V_k^{\rm true}\big|^2}{\sigma_k^2},
\end{equation}
where $\sigma_k^2$ is the thermal noise variance in the visibility measured by baseline $\mathbfit{b}_k$, $V_k^{\rm true}$ is the expected visibility, and $V_k^{\rm cal}$ is the calibrated visibility, which we compute as
\begin{equation}
    \label{eq:cal-data}
    V_k^{\rm cal} = \frac{G_k^{\rm true}}{\hat{G}_k'} \bigl(V_k^{\rm true} + n_k\bigr),
\end{equation}
where $G_k = g_{k_1} g_{k_2}^*$.
Together with the degrees of freedom in the computed $\chi^2$, we may associate each $\chi^2$ measurement with a $p$-value via
\begin{equation}
    \label{eq:p-value}
    p = 1 - F_{\chi^2_{\rm DoF}}(\chi^2),
\end{equation}
where $F_{\chi^2_{\rm DoF}}(x)$ is the cumulative probability of randomly drawing $x$ from a $\chi^2$ distribution with DoF degrees of freedom.
Provided we have correctly counted the degrees of freedom in the $\chi^2$ measurements, we know precisely how the resulting $\chi^2$ values should be distributed if the calibration errors are consistent with the thermal noise injected into the simulations.
Additionally, if the calibration errors are consistent with the injected thermal noise, then the corresponding $p$-value distribution should be uniform on $[0,1]$.
We leverage this knowledge to assess the quality of the calibration by histogramming the measured $\chi^2/{\rm DoF}$ values and their corresponding $p$-values, then comparing the histograms against the expected distributions.

\begin{figure*}
    \includegraphics[width=\textwidth]{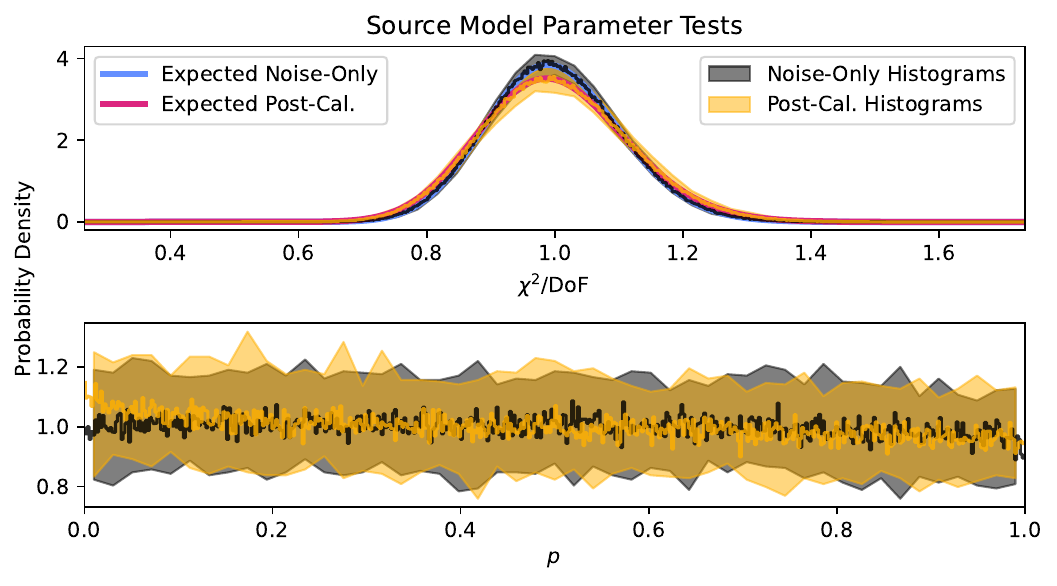}
    \caption{
        $\chi^2/{\rm DoF}$ histograms (\emph{top panel}) and $p$-value histograms (\emph{bottom panel}) for the source model tests described in~\autoref{sec:PTSRC-TESTS}.
        The stepped lines indicate histograms computed using the full set of 720,000 samples, and the shaded regions indicate the range of histogram values obtained when histograms are computed over the 1,000 samples used per unique combination of source flux error, number of sources, and field type.
        The yellow lines and yellow shaded regions correspond to the histograms computed using the calibrated visibilities, while the black lines and black shaded regions correspond to the histograms computed for the ``perfectly calibrated'' visibilities.
        The thick solid lines in the top panel indicate the expected $\chi^2/{\rm DoF}$ distributions when deviations from the true visibilities are consistent with the level of thermal noise injected into the simulations.
        The $\chi^2/{\rm DoF}$ histograms are in excellent agreement with the expected distribution, and the $p$-value histograms are in excellent agreement with a uniform distribution, which suggests that there are no systematic calibration errors associated with missing flux or errors in the modeled source fluxes.
    }
    \label{fig:src-test-results}
\end{figure*}

\autoref{fig:src-test-results} show how the histogrammed $\chi^2/{\rm DoF}$ values and $p$-values compare against the distributions we expect when the residuals are consistent with the injected thermal noise.
Rather than report histograms for individual trials, we opt to report a single histogram taken over all trials.
This decision, which we feel is well motivated because the individual distributions appeared to be consistent with one another under visual inspection, allows us to obtain a more precise estimate of the empirical distribution of calibration errors.
In each panel of~\autoref{fig:src-test-results}, the stepped lines indicate the histogram of the entire set of $\chi^2/{\rm DoF}$ measurements and $p$-values, while the shaded regions indicate the range of probability densities obtained when histogramming the results from individual trials.
The expected $\chi^2/{\rm DoF}$ distribution is plotted in the top panel as a solid line.
We find that the histograms exhibit excellent agreement with the expected $\chi^2/{\rm DoF}$ distributions and excellent agreement with a uniform $p$-value distribution, which indicates that the residuals in the calibrated visibilities are consistent with the thermal noise injected into the simulations.
We therefore conclude that, when the data are perfectly redundant, CorrCal is robust to errors in the source fluxes and the quality of the calibration solutions is fairly insensitive to the number of sources included in the model---there is no evidence of systematic calibration errors due to missing or incorrect flux in the source model, at least within the bounds of the scenarios investigated in this paper.

\subsection{Nonredundancy Tests}
\label{sec:NONRED-TESTS}
To assess the performance of CorrCal in the presence of array nonredundancy, we perform a set of simulations inspired by those performed in~\citet{Sievers:2017}.
For these tests, we use an $8\times8$ rectangular array of parabolic reflector antennas.
The dish of each antenna has a diameter of 14 wavelengths, and nearest neighbors are separated by 20 wavelengths.
We ran 100 simulations for each sky model used in~\autoref{sec:PTSRC-TESTS} for a total of 3000 test simulations.
In each simulation, we apply Gaussian random perturbations to the antenna positions in the plane of the array with a standard deviation of 0.04 wavelengths (or roughly 0.003 dish diameters) in each direction.
This is on the higher end of realistic positioning errors---for example, HERA dishes deviate from a regular grid at the scale of a few centimeters (so slightly less than 0.003 dish diameters), and CHORD dishes are expected to deviate from a regular grid at about the millimeter scale.
We simulate per-antenna gains and initial guesses in the same fashion as described in~\autoref{sec:PTSRC-TESTS}, and add a small amount of radiometer noise to the visibilities (at roughly one fifth the level used in~\autoref{sec:PTSRC-TESTS}).

For each simulation, we perform calibration four different ways, each starting from the same initial guess at the per-antenna gains.
\begin{enumerate}
    \item The first calibration run uses tools from the \texttt{hera\_cal} package\footnote{\url{https://github.com/HERA-Team/hera_cal/}} to perform redundant calibration.
    Since redundant calibration also requires a reasonable initial guess for the redundant visibilities, we provide the redundantly-averaged noiseless visibilities as an initial guess.
    This calibration run serves as our reference, as it is representative of the initial calibration step applied to arrays like HERA~\citep{Dillon:2020}.
    
    \item The second calibration run uses CorrCal in a redundant calibration analog, where no point sources are modeled (i.e., $\Sigma_{kj} = 0$) and the diffuse matrix $\mathbf{\Delta}$ is computed using the ideal antenna positions.
    Since the array is assumed to be perfectly redundant in this scenario, we only use a single complex eigenmode (or two real eigenmodes) for this test.
    This is representative of what one might do when there are unknown positioning errors in the array and a lack of good calibration sources.
    In addition, this test serves as a crucial sanity check, as we should expect to obtain results very similar to those obtained by redundant calibration.
    
    \item The third calibration run uses CorrCal without point source information but uses the exact antenna positions when computing the diffuse matrix.
    This scenario brings us closer to a realistic application of CorrCal, where we have surveyed the actual antenna positions but have not yet brought to bear information about bright sources on the sky.
    
    \item The fourth calibration run uses the exact antenna positions to compute the diffuse matrix and includes the ten point sources that contribute the most to the observed flux in the source matrix $\mathbf{\Sigma}$.
    This is representative of how one would run CorrCal as-is for a real experiment, leveraging information about array irregularities as well as about the sky.
    
\end{enumerate}
For the last two calibration runs, we use the three complex eigenmodes (or six real modes) with the largest eigenvalues for each redundant block when constructing the diffuse matrix.
Loosely speaking, the first eigenmode captures the redundant visibility for each group, while the next two modes capture the covariance structure related to the North-South and East-West antenna position perturbations.
Eigenmodes beyond the first three have values that are suppressed by a factor of roughly $10^8$ relative to the largest eigenmode and a factor of roughly $10^4$ relative to the next two largest eigenmodes.
Just as in~\autoref{sec:PTSRC-TESTS}, we apply an overall amplitude and phase slope correction to the gain solutions obtained from each of the four calibration runs.

\begin{figure*}
    \includegraphics[width=\textwidth]{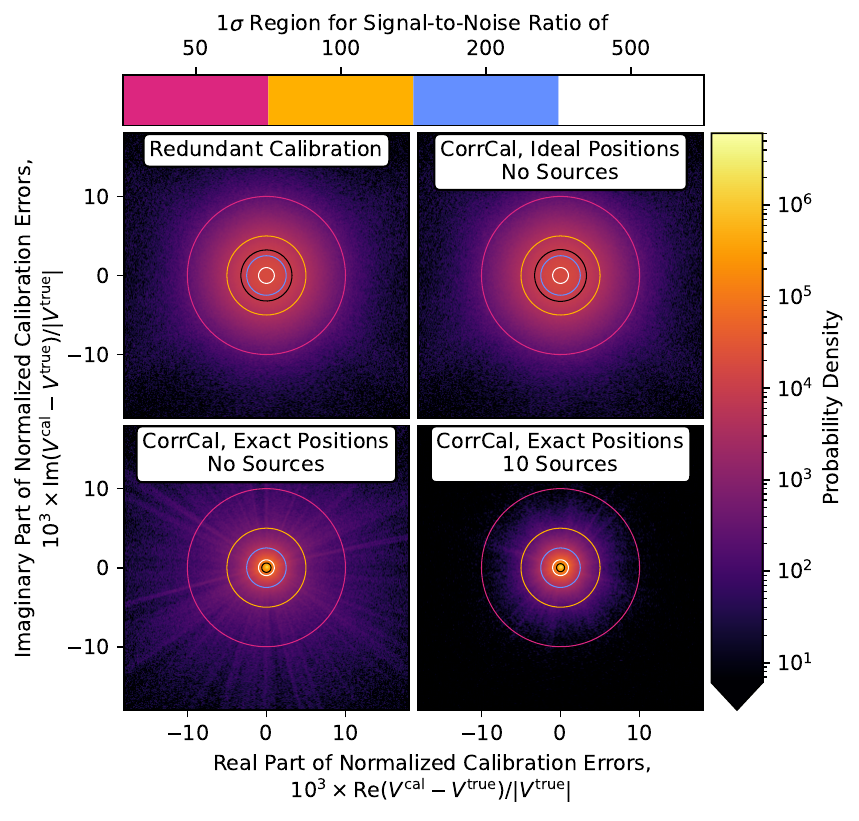}
    \caption{
        Histograms of the calibrated visibility residuals plotted in the complex plane for the four versions of calibration tested in~\autoref{sec:NONRED-TESTS}.
        The calibration errors are computed relative to the true visibilities and normalized by the amplitude of the true visibilities.
        In the upper-left panel, we show the errors from running redundant calibration when the initial guess at the redundant visibilities is the redundant average of the true visibilities.
        In the upper-right panel, we show the errors from running CorrCal using a ``redundant calibration analog'' model covariance that excludes point sources and uses the ideal array layout for computing the diffuse matrix.
        Notably, the upper-right and upper-left panels are visually identical, indicating that applying CorrCal in ``redundant calibration mode'' reproduces the same errors as one gets when applying standard redundant calibration, as expected.
        In the bottom-left panel, we show the errors from running CorrCal where the model covariance excludes point sources and uses the exact antenna positions for computing the diffuse matrix.
        In the bottom-right panel, we show the errors from running CorrCal where the model covariance uses the exact antenna positions for computing both the source matrix and the diffuse matrix, where the ten sources that contribute the most to the observed flux are included in the source model.
        In each panel we also plot four contours corresponding to the $1\sigma$ region for a complex Gaussian distribution with a given signal-to-noise ratio, as well as a black contour indicating the regions containing approximately 68\% of the samples in the histograms.
    }
    \label{fig:nonredundancy-results}
\end{figure*}

In~\autoref{fig:nonredundancy-results}, we show histograms of the residuals between the calibrated data and the true data, normalized to the amplitude of the true data.
Concretely, we compute
\begin{equation}
    r_k = \frac{\alpha_k \bigl(V_k^{\rm true} + n_k\bigr) - V_k^{\rm true}}{\big|V_k^{\rm true}\big|},
\end{equation}
where $\alpha_k \equiv G_k^{\rm true}/\hat{G}_k'$, and histogram $r_k$ in the complex plane.
We choose to histogram this quantity because it allows us to directly compare the calibration errors against thermal noise fluctuations.
In the limit of a perfect calibration, $r_k$ is just the noise normalized to the visibility amplitude, while in the high signal-to-noise limit (which applies to the tests considered here), the variability in $r_k$ is driven by systematic errors in the calibration.
We therefore choose to plot on top of the histograms four contours corresponding to the $1\sigma$ region for normalized noise fluctuations of varying signal-to-noise ratios.
In addition to the four colored contours, we plot in black the contours that contain approximately 68\% of the samples in each histogram.

There is a clear hierarchy in the results shown in~\autoref{fig:nonredundancy-results}.
Redundant calibration and the redundant analog version of CorrCal perform comparably to one another and produce the greatest errors in the calibrated visibilities of the four methods tested.
For these first two cases, the fractional errors are mostly contained to 1\% or less, with a shallow peak near zero and a rapid decrease in probability density beyond the 1\% level.
Including exact antenna positions when modeling the diffuse matrix significantly reduces the calibration errors, as evidenced by the more compact error distribution (which instead falls off around the 0.5\% level) and the sharper peak near zero.\footnote{The radial spurs in the error distribution are a consequence of excluding information about bright sources---although not shown here, these features vanish when the error distribution is computed without the results from the fields containing a very bright point source.}
Including information about point sources and accurate antenna positions provides the best results: the corresponding error distribution shows the sharpest peak near zero and the probability density of the fractional errors sharply falls off beyond the 0.5\% level.

The salient interpretation of the results shown in~\autoref{fig:nonredundancy-results} is that the calibration errors associated with nonredundancy in the data are unbiased and subdominant to thermal fluctuations in the visibilities at the typical single integration and single channel level.
For many modern drift-scanning interferometers, the signal-to-noise ratio in the cross-correlations is typically not very high.
For example, in the simulations run in~\autoref{sec:REALISTIC-SKY}, the radiometer noise associated with a 10\,s integration and 100\,kHz channel width produces a signal-to-noise ratio that is on average less than 10.
We therefore conclude that modest levels of nonredundancy will not significantly impact the quality of calibration solutions in realistic applications of CorrCal.
We temper this conclusion, however, by noting that the calibration errors due to nonredundancy might not integrate down with time and may present as a systematic error in highly averaged data.

In addition to investigating the distribution of errors in the calibrated visibilities, we analyze the errors in the gain solutions themselves.
To obtain a sense of how stable the gain solutions are across the array, we compute the standard deviation in the gain amplitude errors, $|g^{\rm true}| - |\hat{g}^\prime|$, as well as the standard deviation in the gain phase errors, $\phi^{\rm true} - \phi^\prime ={\rm arg}(g^{\rm true} \hat{g}^{\prime*})$, across the array.
In~\autoref{table:nonred-results}, we provide a summary of our results by indicating the bounds containing 95\% of the amplitude and phase error standard deviations.
The results in~\autoref{table:nonred-results} provide an alternate insight into the hierarchy displayed in~\autoref{fig:nonredundancy-results}: as more information is added to the calibration model, there is less variance in the calibration errors.

\begin{deluxetable}{ccc}
    \tablehead{\colhead{Calibration Type} & \colhead{$\sigma\bigl(|g^{\rm true}| - |\hat{g}^\prime|\bigr)$} & \colhead{$\sigma\bigl(\phi^{\rm true} - \hat{\phi}^\prime\bigr)$}}
    \tablewidth{\columnwidth}
    \tablecaption{
        \label{table:nonred-results} Summary of Errors in Gain Solutions
    }
    \startdata
        Redundant Calibration & $2.5^{+0.8}_{-0.8} \times 10^{-3}$ & $2.7^{+8.8}_{-1.5} \times 10^{-3}$ \\
        CorrCal, Ideal Positions & $2.6^{+0.8}_{-0.9} \times 10^{-3}$ & $2.7^{+8.8}_{-1.5} \times 10^{-3}$ \\
        CorrCal, Exact Positions & $3.1^{+4.9}_{-1.5} \times 10^{-5}$ & $1.9^{+7.8}_{-1.8} \times 10^{-3}$  \\
        CorrCal, with Sources & $2.4^{+1.3}_{-1.2} \times 10^{-5}$ & $0.8^{+1.1}_{-0.6}\times 10^{-3}$ 
    \enddata
\end{deluxetable}

\subsection{Non-Gaussian Sky Test}
\label{sec:REALISTIC-SKY}
As a final test, we ran CorrCal on simulated data using a realistic model of the sky when simulating visibilities.
Importantly, this test explicitly breaks the assumption of Gaussianity in the diffuse emission on the sky, essentially bringing to bear an end-to-end test of the claim made in~\autoref{sec:MODEL-COVARIANCE} and in~\citet{Sievers:2017} that non-Gaussianity in the foregrounds would not affect the quality of the calibration solutions.
We simulate data for a 61-element hexagonal array located at the HERA site at a local sidereal time of $2^h 0^m$ and a frequency of 151\,MHz.
Each antenna is 14\,m in diameter with a diffraction-limited (i.e., Airy disk) beam and nearest neighbor antennas are separated by 14.6\,m.
The synthesized beam for this array has a full-width at half-maximum of about one degree.
The point sources in the sky model are taken from the GLEAM catalog~\citep{Hurley-Walker:2017} using the integrated flux densities at 151\,MHz and including all sources above the horizon (roughly 161,000 sources in total) at the chosen observation time and telescope location.
We additionally include the Global Sky Model~\citep{deOliveira-Costa:2008} interpolated to 151\,MHz and downsampled to a resolution of roughly half a degree as the diffuse sky component.
We simulate gains in the same manner as the previous sections but do not add any thermal noise to the simulated visibilities.
For the source matrix, we include the top ten contributors to the observed flux, and for the diffuse matrix we use the angular power spectrum of the interpolated Global Sky Model.

\begin{figure}
    \includegraphics[width=\columnwidth]{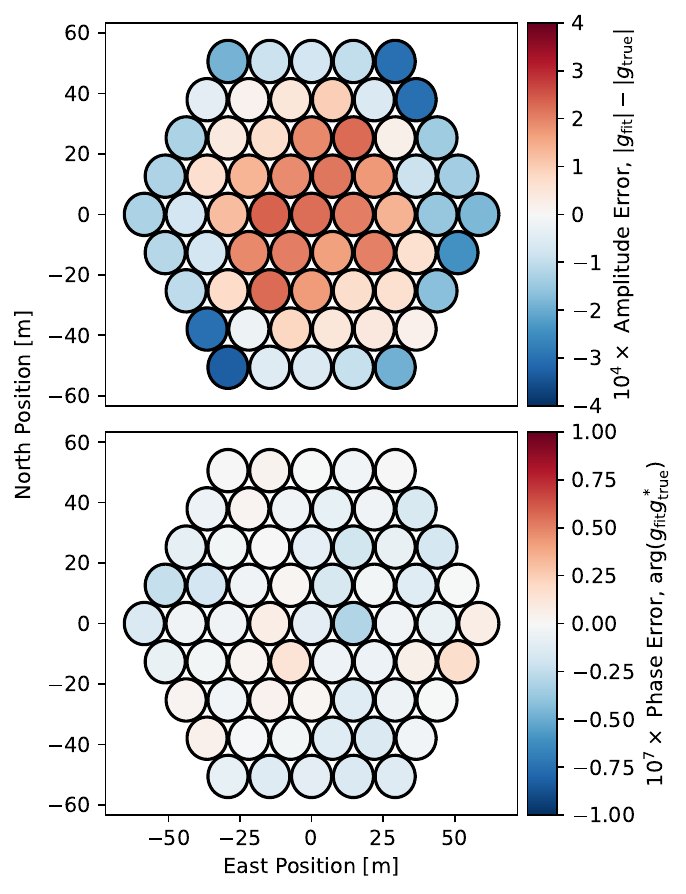}
    \caption{
        Errors in the gain amplitudes (\emph{top panel}) and the gain phases (\emph{bottom panel}) obtained for noiseless simulations using a realistic sky, as described in~\autoref{sec:REALISTIC-SKY}.
        The gain errors are computed after applying an overall amplitude and phase slope correction to the calibration solutions.
        Errors in the gain amplitudes and gain phases are small, which suggests that non-Gaussianities in the diffuse emission do not strongly impact the quality of the gain solutions.
    }
    \label{fig:realistic-sky-results}
\end{figure}

In~\autoref{fig:realistic-sky-results} we show the errors in the amplitude and phase of the gain solutions.
As in the previous sections, prior to computing the residuals we apply an overall amplitude correction and a phase slope correction to the calibration solution.
The errors in the calibration solutions are small: amplitude errors typically occur at the $\lesssim 10^{-4}$ level and phase errors are typically around $10^{-7}$\,rad, however the distribution of amplitude and phase errors may change depending on which part of the sky is overhead.
While this is not an exhaustive test of how CorrCal performs with a realistic sky, these results suggest that calibration errors due to non-Gaussianities in the diffuse emission are negligible in applications with realistic noise levels.
We therefore conclude that treating the diffuse emission on the sky as a Gaussian random field is valid for the purposes of calibration via CorrCal.

\section{Conclusion}
\label{sec:CONCLUSION}
In this paper, we provided an overview of the formalism supporting CorrCal and presented the results of a set of tests designed to assess the accuracy of CorrCal in the presence of various modeling errors.
We showed that, under an appropriate set of assumptions, the covariance between visibilities takes on a sparse form which may be leveraged to efficiently perform a covariance-based calibration of radio interferometric data.Slide 52
We found that CorrCal can obtain accurate calibration solutions by relying only on a handful of point sources, a model of the primary beam, and the power spectrum of the diffuse emission on the sky.
Moreover, the accuracy of the calibration solutions is relatively insensitive to missing or inaccurate source flux densities and positional nonredundancy in the array layout.
A limited test with a realistic sky model also revealed that non-Gaussianities in the diffuse emission do not strongly affect the quality of the calibration solutions obtained with CorrCal.
These tests collectively provide strong foundational support for CorrCal's ability to obtain high quality calibration solutions in the presence of various modeling errors and array imperfections.

Although we have performed an extensive set of foundational tests of CorrCal, there is still a broad scope of important tests of CorrCal and extensions to CorrCal that remain to be implemented.
Future work investigating CorrCal's resilience to modeling errors will look into how other forms of nonredundancy, such as beam-to-beam variations or mutual coupling between array elements, affect the quality of calibration solutions.
A crucially important investigation that is missing from this initial analysis is a study of whether CorrCal injects spectral structure into the calibrated visibilities when different frequency channels are calibrated independently of one another.
To that end, work is underway on building an extension of CorrCal that uses a model of correlations between frequency channels to fit the full bandpass for each antenna.
Other natural extensions include a polarization-aware version of CorrCal, the addition of new calibration parameters to fit for antenna-based variations (e.g., pointing errors or parametrized beam corrections), and a GPU-based implementation to enable real-time calibration for next-generation arrays.
To close out this non-exhaustive list, another important question to be answered is whether CorrCal performs better under a different assumption about the statistics of the data.

Given the demonstrated success of CorrCal in our suite of validation tests and the broad range of avenues for further exploration, we are enthusiastic about future developments to CorrCal and forthcoming results of CorrCal applied to data from current and next-generation experiments, such as HERA, CHORD, and HIRAX.

\vspace{2\baselineskip}
\section*{Acknowledgements}
The authors thank Miguel Morales, Michael Wilensky, Kendrick Smith, Leon Koopmans, and Ruby Byrne for insightful conversations and helpful feedback.
A.L. acknowledges support from an NSERC Discovery Grant, an Alliance International Grant, and the William Dawson Scholarship at McGill.
R.P. acknowledges support from the Faculty of Arts \& Science at the University of Toronto and the Dunlap Institute.
The Dunlap Institute is funded through an endowment established by the David Dunlap family and the University of Toronto.
This work was funded in part by the Canada 150 Research Chairs Program.
Research at Perimeter Institute is supported in part by the Government of Canada through the Department of Innovation, Science and Economic Development Canada and by the Province of Ontario through the Ministry of Colleges and Universities.

\software{
    numpy~\citep{numpy},
    scipy~\citep{scipy},
    matplotlib~\citep{matplotlib},
    astropy~\citep{astropy},
    pyuvdata~\citep{pyuvdata}
}

\bibliography{references}{}

@article{Byrne:2019,
doi = {10.3847/1538-4357/ab107d},
url = {https://dx.doi.org/10.3847/1538-4357/ab107d},
year = {2019},
month = {apr},
publisher = {The American Astronomical Society},
volume = {875},
number = {1},
pages = {70},
author = {Ruby Byrne and Miguel F. Morales and Bryna Hazelton and Wenyang Li and Nichole Barry and Adam P. Beardsley and Ronniy Joseph and Jonathan Pober and Ian Sullivan and Cathryn Trott},
title = {Fundamental Limitations on the Calibration of Redundant 21 cm Cosmology Instruments and Implications for HERA and the SKA},
journal = {The Astrophysical Journal}
}

@article{Barry:2016,
    author = {Barry, N. and Hazelton, B. and Sullivan, I. and Morales, M. F. and Pober, J. C.},
    title = "{Calibration requirements for detecting the 21 cm epoch of reionization power spectrum and implications for the SKA}",
    journal = {Monthly Notices of the Royal Astronomical Society},
    volume = {461},
    number = {3},
    pages = {3135-3144},
    year = {2016},
    month = {06},
    issn = {0035-8711},
    doi = {10.1093/mnras/stw1380},
    url = {https://doi.org/10.1093/mnras/stw1380},
    eprint = {https://academic.oup.com/mnras/article-pdf/461/3/3135/8106921/stw1380.pdf},
}

@article{Orosz:2019,
    author = {Orosz, Naomi and Dillon, Joshua S and Ewall-Wice, Aaron and Parsons, Aaron R and Thyagarajan, Nithyanandan},
    title = "{Mitigating the effects of antenna-to-antenna variation on redundant-baseline calibration for 21 cm cosmology}",
    journal = {Monthly Notices of the Royal Astronomical Society},
    volume = {487},
    number = {1},
    pages = {537-549},
    year = {2019},
    month = {05},
    issn = {0035-8711},
    doi = {10.1093/mnras/stz1287},
    url = {https://doi.org/10.1093/mnras/stz1287},
    eprint = {https://academic.oup.com/mnras/article-pdf/487/1/537/28705585/stz1287.pdf},
}

@article{Charles:2024,
    author = {Charles, N and Kern, N S and Pascua, R and Bernardi, G and Bester, L and Smirnov, O and Acedo, E D L and Abdurashidova, Z and Adams, T and Aguirre, J E and Baartman, R and Beardsley, A P and Berkhout, L M and Billings, T S and Bowman, J D and Bull, P and Burba, J and Byrne, R and Carey, S and Chen, K and Choudhuri, S and Cox, T and DeBoer, D R and Dexter, M and Dillon, J S and Dynes, S and Eksteen, N and Ely, J and Ewall-Wice, A and Fritz, R and Furlanetto, S R and Gale-Sides, K and Garsden, H and Gehlot, B K and Ghosh, A and Gorce, A and Gorthi, D and Halday, Z and Hazelton, B J and Hewitt, J N and Hickish, J and Huang, T and Jacobs, D C and Josaitis, A and Kerrigan, J and Kittiwisit, P and Kolopanis, M and Lanman, A and Liu, A and Ma, Y -Z and MacMahon, D H E and Malan, L and Malgas, K and Malgas, C and Marero, B and Martinot, Z E and McBride, L and Mesinger, A and Mohamed-Hinds, N and Molewa, M and Morales, M F and Murray, S and Nikolic, B and Nuwegeld, H and Parsons, A R and Patra, N and Plante, P L and Qin, Y and Rath, E and Razavi-Ghods, N and Riley, D and Robnett, J and Rosie, K and Santos, M G and Sims, P and Singh, S and Storer, D and Swarts, H and Tan, J and Wilensky, M J and Williams, P K G and Wyngaarden, P v and Zheng, H},
    title = {Mitigating calibration errors from mutual coupling with time-domain filtering of 21 cm cosmological radio observations},
    journal = {Monthly Notices of the Royal Astronomical Society},
    volume = {534},
    number = {4},
    pages = {3349-3363},
    year = {2024},
    month = {10},
    issn = {0035-8711},
    doi = {10.1093/mnras/stae2303},
    url = {https://doi.org/10.1093/mnras/stae2303},
    eprint = {https://academic.oup.com/mnras/article-pdf/534/4/3349/59933025/stae2303.pdf},
}

@article{Byrne:2021,
    author = {Byrne, Ruby and Morales, Miguel F and Hazelton, Bryna J and Wilensky, Michael},
    title = {A unified calibration framework for 21 cm cosmology},
    journal = {Monthly Notices of the Royal Astronomical Society},
    volume = {503},
    number = {2},
    pages = {2457-2477},
    year = {2021},
    month = {03},
    issn = {0035-8711},
    doi = {10.1093/mnras/stab647},
    url = {https://doi.org/10.1093/mnras/stab647},
    eprint = {https://academic.oup.com/mnras/article-pdf/503/2/2457/36686004/stab647.pdf},
}

@article{Byrne:2023,
doi = {10.3847/1538-4357/acac95},
url = {https://dx.doi.org/10.3847/1538-4357/acac95},
year = {2023},
month = {feb},
publisher = {The American Astronomical Society},
volume = {943},
number = {2},
pages = {117},
author = {Byrne, Ruby},
title = {Delay-weighted Calibration: Precision Calibration for 21 cm Cosmology with Resilience to Sky Model Error},
journal = {The Astrophysical Journal}
}

@article{Yatawatta:2009,
  title={IEEE 13th Digital Signal Processing Workshop and 5th IEEE Signal Processing Education Workshop},
  author={Yatawatta, S and Zaroubi, S and de Bruyn, G and Koopmans, L and Noordam, J},
  journal={IEEE, Piscataway, NJ},
  pages={150},
  year={2009}
}

@article{Ewall-Wice:2022b,
doi = {10.3847/1538-4357/ac87b3},
url = {https://dx.doi.org/10.3847/1538-4357/ac87b3},
year = {2022},
month = {oct},
publisher = {The American Astronomical Society},
volume = {938},
number = {2},
pages = {151},
author = {Ewall-Wice, Aaron and Dillon, Joshua S. and Gehlot, Bharat and Parsons, Aaron and Cox, Tyler and Jacobs, Daniel C.},
title = {Precision Calibration of Radio Interferometers for 21 cm Cosmology with No Redundancy and Little Knowledge of Antenna Beams and the Radio Sky},
journal = {The Astrophysical Journal}
}

@article{Liu:2010,
    author = {Liu, Adrian and Tegmark, Max and Morrison, Scott and Lutomirski, Andrew and Zaldarriaga, Matias},
    title = {Precision calibration of radio interferometers using redundant baselines},
    journal = {Monthly Notices of the Royal Astronomical Society},
    volume = {408},
    number = {2},
    pages = {1029-1050},
    year = {2010},
    month = {07},
    issn = {0035-8711},
    doi = {10.1111/j.1365-2966.2010.17174.x},
    url = {https://doi.org/10.1111/j.1365-2966.2010.17174.x},
    eprint = {https://academic.oup.com/mnras/article-pdf/408/2/1029/18440559/mnras0408-1029.pdf},
}

@article{Dillon:2018,
    author = {Dillon, Joshua S and Kohn, Saul A and Parsons, Aaron R and Aguirre, James E and Ali, Zaki S and Bernardi, Gianni and Kern, Nicholas S and Li, Wenyang and Liu, Adrian and Nunhokee, Chuneeta D and Pober, Jonathan C},
    title = {Polarized redundant-baseline calibration for 21 cm cosmology without adding spectral structure},
    journal = {Monthly Notices of the Royal Astronomical Society},
    volume = {477},
    number = {4},
    pages = {5670-5681},
    year = {2018},
    month = {04},
    issn = {0035-8711},
    doi = {10.1093/mnras/sty1060},
    url = {https://doi.org/10.1093/mnras/sty1060},
    eprint = {https://academic.oup.com/mnras/article-pdf/477/4/5670/24955748/sty1060.pdf},
}

@ARTICLE{Joseph:2018,
       author = {{Joseph}, Ronniy C. and {Trott}, Cathryn M. and {Wayth}, Randall B.},
        title = "{The Bias and Uncertainty of Redundant and Sky-based Calibration Under Realistic Sky and Telescope Conditions}",
      journal = {\aj},
         year = 2018,
        month = dec,
       volume = {156},
       number = {6},
          eid = {285},
        pages = {285},
          doi = {10.3847/1538-3881/aaec0b},
archivePrefix = {arXiv},
       eprint = {1810.11237},
 primaryClass = {astro-ph.IM},
       adsurl = {https://ui.adsabs.harvard.edu/abs/2018AJ....156..285J}
}

@ARTICLE{Li:2018,
       author = {{Li}, W. and {Pober}, J.~C. and {Hazelton}, B.~J. and {Barry}, N. and {Morales}, M.~F. and {Sullivan}, I. and {Parsons}, A.~R. and {Ali}, Z.~S. and {Dillon}, J.~S. and {Beardsley}, A.~P. and {Bowman}, J.~D. and {Briggs}, F. and {Byrne}, R. and {Carroll}, P. and {Crosse}, B. and {Emrich}, D. and {Ewall-Wice}, A. and {Feng}, L. and {Franzen}, T.~M.~O. and {Hewitt}, J.~N. and {Horsley}, L. and {Jacobs}, D.~C. and {Johnston-Hollitt}, M. and {Jordan}, C. and {Joseph}, R.~C. and {Kaplan}, D.~L. and {Kenney}, D. and {Kim}, H. and {Kittiwisit}, P. and {Lanman}, A. and {Line}, J. and {McKinley}, B. and {Mitchell}, D.~A. and {Murray}, S. and {Neben}, A. and {Offringa}, A.~R. and {Pallot}, D. and {Paul}, S. and {Pindor}, B. and {Procopio}, P. and {Rahimi}, M. and {Riding}, J. and {Sethi}, S.~K. and {Udaya Shankar}, N. and {Steele}, K. and {Subrahmanian}, R. and {Tegmark}, M. and {Thyagarajan}, N. and {Tingay}, S.~J. and {Trott}, C. and {Walker}, M. and {Wayth}, R.~B. and {Webster}, R.~L. and {Williams}, A. and {Wu}, C. and {Wyithe}, S.},
        title = "{Comparing Redundant and Sky-model-based Interferometric Calibration: A First Look with Phase II of the MWA}",
      journal = {\apj},
         year = 2018,
        month = aug,
       volume = {863},
       number = {2},
          eid = {170},
        pages = {170},
          doi = {10.3847/1538-4357/aad3c3},
archivePrefix = {arXiv},
       eprint = {1807.05312},
 primaryClass = {astro-ph.IM},
       adsurl = {https://ui.adsabs.harvard.edu/abs/2018ApJ...863..170L}
}

@article{Kern:2020b,
doi = {10.3847/1538-4357/ab67bc},
url = {https://dx.doi.org/10.3847/1538-4357/ab67bc},
year = {2020},
month = {feb},
publisher = {The American Astronomical Society},
volume = {890},
number = {2},
pages = {122},
author = {Kern, Nicholas S. and Dillon, Joshua S. and Parsons, Aaron R. and Carilli, Christopher L. and Bernardi, Gianni and Abdurashidova, Zara and Aguirre, James E. and Alexander, Paul and Ali, Zaki S. and Balfour, Yanga and Beardsley, Adam P. and Billings, Tashalee S. and Bowman, Judd D. and Bradley, Richard F. and Bull, Philip and Burba, Jacob and Carey, Steven and Cheng, Carina and DeBoer, David R. and Dexter, Matt and de Lera Acedo, Eloy and Ely, John and Ewall-Wice, Aaron and Fagnoni, Nicolas and Fritz, Randall and Furlanetto, Steve R. and Gale-Sides, Kingsley and Glendenning, Brian and Gorthi, Deepthi and Greig, Bradley and Grobbelaar, Jasper and Halday, Ziyaad and Hazelton, Bryna J. and Hewitt, Jacqueline N. and Hickish, Jack and Jacobs, Daniel C. and Julius, Austin and Kerrigan, Joshua and Kittiwisit, Piyanat and Kohn, Saul A. and Kolopanis, Matthew and Lanman, Adam and La Plante, Paul and Lekalake, Telalo and Liu, Adrian and MacMahon, David and Malan, Lourence and Malgas, Cresshim and Maree, Matthys and Martinot, Zachary E. and Matsetela, Eunice and Mesinger, Andrei and Molewa, Mathakane and Morales, Miguel F. and Mosiane, Tshegofalang and Murray, Steven G. and Neben, Abraham R. and Nikolic, Bojan and Nunhokee, Chuneeta D. and Patra, Nipanjana and Pieterse, Samantha and Pober, Jonathan C. and Razavi-Ghods, Nima and Ringuette, Jon and Robnett, James and Rosie, Kathryn and Sims, Peter and Smith, Craig and Syce, Angelo and Thyagarajan, Nithyanandan and Williams, Peter K. G. and Zheng, Haoxuan},
title = {Absolute Calibration Strategies for the Hydrogen Epoch of Reionization Array and Their Impact on the 21 cm Power Spectrum},
journal = {The Astrophysical Journal}
}

@misc{Sievers:2017,
      title={Calibration of Quasi-Redundant Interferometers}, 
      author={J. L. Sievers},
      year={2017},
      eprint={1701.01860},
      archivePrefix={arXiv},
      primaryClass={astro-ph.IM},
      url={https://arxiv.org/abs/1701.01860}, 
}

@article{Dillon:2020,
    author = {Dillon, Joshua S and Lee, Max and Ali, Zaki S and Parsons, Aaron R and Orosz, Naomi and Nunhokee, Chuneeta Devi and La Plante, Paul and Beardsley, Adam P and Kern, Nicholas S and Abdurashidova, Zara and Aguirre, James E and Alexander, Paul and Balfour, Yanga and Bernardi, Gianni and Billings, Tashalee S and Bowman, Judd D and Bradley, Richard F and Bull, Phil and Burba, Jacob and Carey, Steve and Carilli, Chris L and Cheng, Carina and DeBoer, David R and Dexter, Matt and de Lera Acedo, Eloy and Ely, John and Ewall-Wice, Aaron and Fagnoni, Nicolas and Fritz, Randall and Furlanetto, Steven R and Gale-Sides, Kingsley and Glendenning, Brian and Gorthi, Deepthi and Greig, Bradley and Grobbelaar, Jasper and Halday, Ziyaad and Hazelton, Bryna J and Hewitt, Jacqueline N and Hickish, Jack and Jacobs, Daniel C and Julius, Austin and Kerrigan, Joshua and Kittiwisit, Piyanat and Kohn, Saul A and Kolopanis, Matthew and Lanman, Adam and Lekalake, Telalo and Lewis, David and Liu, Adrian and Ma, Yin-Zhe and MacMahon, David and Malan, Lourence and Malgas, Cresshim and Maree, Matthys and Martinot, Zachary E and Matsetela, Eunice and Mesinger, Andrei and Molewa, Mathakane and Morales, Miguel F and Mosiane, Tshegofalang and Murray, Steven and Neben, Abraham R and Nikolic, Bojan and Pascua, Robert and Patra, Nipanjana and Pieterse, Samantha and Pober, Jonathan C and Razavi-Ghods, Nima and Ringuette, Jon and Robnett, James and Rosie, Kathryn and Santos, Mario G and Sims, Peter and Smith, Craig and Syce, Angelo and Tegmark, Max and Thyagarajan, Nithyanandan and Williams, Peter K G and Zheng, Haoxuan},
    title = {Redundant-baseline calibration of the hydrogen epoch of reionization array},
    journal = {Monthly Notices of the Royal Astronomical Society},
    volume = {499},
    number = {4},
    pages = {5840-5861},
    year = {2020},
    month = {10},
    issn = {0035-8711},
    doi = {10.1093/mnras/staa3001},
    url = {https://doi.org/10.1093/mnras/staa3001},
    eprint = {https://academic.oup.com/mnras/article-pdf/499/4/5840/34158004/staa3001.pdf},
}

@article{Cox:2024,
    author = {Cox, Tyler A and Parsons, Aaron R and Dillon, Joshua S and Ewall-Wice, Aaron and Pascua, Robert},
    title = {Spectral redundancy for calibrating interferometers and suppressing the foreground wedge in 21 cm cosmology},
    journal = {Monthly Notices of the Royal Astronomical Society},
    volume = {532},
    number = {3},
    pages = {3375-3394},
    year = {2024},
    month = {07},
    issn = {0035-8711},
    doi = {10.1093/mnras/stae1612},
    url = {https://doi.org/10.1093/mnras/stae1612},
    eprint = {https://academic.oup.com/mnras/article-pdf/532/3/3375/58633806/stae1612.pdf},
}

@article{Sims:2022a,
    author = {Sims, Peter H and Pober, Jonathan C and Sievers, Jonathan L},
    title = {A Bayesian approach to high-fidelity interferometric calibration – I. Mathematical formalism},
    journal = {Monthly Notices of the Royal Astronomical Society},
    volume = {517},
    number = {1},
    pages = {910-934},
    year = {2022},
    month = {07},
    issn = {0035-8711},
    doi = {10.1093/mnras/stac1861},
    url = {https://doi.org/10.1093/mnras/stac1861},
    eprint = {https://academic.oup.com/mnras/article-pdf/517/1/910/46395178/stac1861.pdf},
}

@article{Sims:2022b,
    author = {Sims, Peter H and Pober, Jonathan C and Sievers, Jonathan L},
    title = {A Bayesian approach to high fidelity interferometric calibration − II: demonstration with simulated data},
    journal = {Monthly Notices of the Royal Astronomical Society},
    volume = {517},
    number = {1},
    pages = {935-961},
    year = {2022},
    month = {07},
    issn = {0035-8711},
    doi = {10.1093/mnras/stac1749},
    url = {https://doi.org/10.1093/mnras/stac1749},
    eprint = {https://academic.oup.com/mnras/article-pdf/517/1/935/46395205/stac1749.pdf},
}

@article{deOliveira-Costa:2008,
    author = {De Oliveira-Costa, Angélica and Tegmark, Max and Gaensler, B. M. and Jonas, Justin and Landecker, T. L. and Reich, Patricia},
    title = {A model of diffuse Galactic radio emission from 10 MHz to 100 GHz},
    journal = {Monthly Notices of the Royal Astronomical Society},
    volume = {388},
    number = {1},
    pages = {247-260},
    year = {2008},
    month = {07},
    issn = {0035-8711},
    doi = {10.1111/j.1365-2966.2008.13376.x},
    url = {https://doi.org/10.1111/j.1365-2966.2008.13376.x},
    eprint = {https://academic.oup.com/mnras/article-pdf/388/1/247/18721410/mnras0388-0247.pdf},
}

@article{Hurley-Walker:2017,
    author = {Hurley-Walker, N and Callingham, J R and Hancock, P J and Franzen, T M O and Hindson, L and Kapińska, A D and Morgan, J and Offringa, A R and Wayth, R B and Wu, C and Zheng, Q and Murphy, T and Bell, M E and Dwarakanath, K S and For, B and Gaensler, B M and Johnston-Hollitt, M and Lenc, E and Procopio, P and Staveley-Smith, L and Ekers, R and Bowman, J D and Briggs, F and Cappallo, R J and Deshpande, A A and Greenhill, L and Hazelton, B J and Kaplan, D L and Lonsdale, C J and McWhirter, S R and Mitchell, D A and Morales, M F and Morgan, E and Oberoi, D and Ord, S M and Prabu, T and Shankar, N Udaya and Srivani, K S and Subrahmanyan, R and Tingay, S J and Webster, R L and Williams, A and Williams, C L},
    title = "{GaLactic and Extragalactic All-sky Murchison Widefield Array (GLEAM) survey – I. A low-frequency extragalactic catalogue}",
    journal = {Monthly Notices of the Royal Astronomical Society},
    volume = {464},
    number = {1},
    pages = {1146-1167},
    year = {2016},
    month = {09},
    issn = {0035-8711},
    doi = {10.1093/mnras/stw2337},
    url = {https://doi.org/10.1093/mnras/stw2337},
    eprint = {https://academic.oup.com/mnras/article-pdf/464/1/1146/47736764/mnras\_464\_1\_1146.pdf},
}

@article{Franzen:2019,
    title={Source counts and confusion at 72–231 MHz in the MWA GLEAM survey},
    volume={36},
    DOI={10.1017/pasa.2018.52},
    journal={Publications of the Astronomical Society of Australia},
    author={Franzen, T. M. O. and Vernstrom, T. and Jackson, C. A. and Hurley-Walker, N. and Ekers, R. D. and Heald, G. and Seymour, N. and White, S. V.},
    year={2019},
    pages={e004}
}

@article{pyuvdata,
  doi = {10.21105/joss.00140},
  url = {https://doi.org/10.21105/joss.00140},
  year = {2017},
  publisher = {The Open Journal},
  volume = {2},
  number = {10},
  pages = {140},
  author = {Bryna J. Hazelton and Daniel C. Jacobs and Jonathan C. Pober and Adam P. Beardsley},
  title = {pyuvdata: an interface for astronomical interferometeric datasets in python},
  journal = {Journal of Open Source Software}
}

@ARTICLE{scipy,
  author  = {Virtanen, Pauli and Gommers, Ralf and Oliphant, Travis E. and
            Haberland, Matt and Reddy, Tyler and Cournapeau, David and
            Burovski, Evgeni and Peterson, Pearu and Weckesser, Warren and
            Bright, Jonathan and {van der Walt}, St{\'e}fan J. and
            Brett, Matthew and Wilson, Joshua and Millman, K. Jarrod and
            Mayorov, Nikolay and Nelson, Andrew R. J. and Jones, Eric and
            Kern, Robert and Larson, Eric and Carey, C J and
            Polat, {\.I}lhan and Feng, Yu and Moore, Eric W. and
            {VanderPlas}, Jake and Laxalde, Denis and Perktold, Josef and
            Cimrman, Robert and Henriksen, Ian and Quintero, E. A. and
            Harris, Charles R. and Archibald, Anne M. and
            Ribeiro, Ant{\^o}nio H. and Pedregosa, Fabian and
            {van Mulbregt}, Paul and {SciPy 1.0 Contributors}},
  title   = {{{SciPy} 1.0: Fundamental Algorithms for Scientific
            Computing in Python}},
  journal = {Nature Methods},
  year    = {2020},
  volume  = {17},
  pages   = {261--272},
  adsurl  = {https://rdcu.be/b08Wh},
  doi     = {10.1038/s41592-019-0686-2},
}

@Article{numpy,
 title         = {Array programming with {NumPy}},
 author        = {Charles R. Harris and K. Jarrod Millman and St{\'{e}}fan J.
                 van der Walt and Ralf Gommers and Pauli Virtanen and David
                 Cournapeau and Eric Wieser and Julian Taylor and Sebastian
                 Berg and Nathaniel J. Smith and Robert Kern and Matti Picus
                 and Stephan Hoyer and Marten H. van Kerkwijk and Matthew
                 Brett and Allan Haldane and Jaime Fern{\'{a}}ndez del
                 R{\'{i}}o and Mark Wiebe and Pearu Peterson and Pierre
                 G{\'{e}}rard-Marchant and Kevin Sheppard and Tyler Reddy and
                 Warren Weckesser and Hameer Abbasi and Christoph Gohlke and
                 Travis E. Oliphant},
 year          = {2020},
 month         = sep,
 journal       = {Nature},
 volume        = {585},
 number        = {7825},
 pages         = {357--362},
 doi           = {10.1038/s41586-020-2649-2},
 publisher     = {Springer Science and Business Media {LLC}},
 url           = {https://doi.org/10.1038/s41586-020-2649-2}
}

@Article{matplotlib,
  Author    = {Hunter, J. D.},
  Title     = {Matplotlib: A 2D graphics environment},
  Journal   = {Computing in Science \& Engineering},
  Volume    = {9},
  Number    = {3},
  Pages     = {90--95},
  publisher = {IEEE COMPUTER SOC},
  doi       = {10.1109/MCSE.2007.55},
  year      = 2007
}

@ARTICLE{astropy,
       author = {{Astropy Collaboration} and {Price-Whelan}, Adrian M. and {Lim}, Pey Lian and {Earl}, Nicholas and {Starkman}, Nathaniel and {Bradley}, Larry and {Shupe}, David L. and {Patil}, Aarya A. and {Corrales}, Lia and {Brasseur}, C.~E. and {N{\"o}the}, Maximilian and {Donath}, Axel and {Tollerud}, Erik and {Morris}, Brett M. and {Ginsburg}, Adam and {Vaher}, Eero and {Weaver}, Benjamin A. and {Tocknell}, James and {Jamieson}, William and {van Kerkwijk}, Marten H. and {Robitaille}, Thomas P. and {Merry}, Bruce and {Bachetti}, Matteo and {G{\"u}nther}, H. Moritz and {Aldcroft}, Thomas L. and {Alvarado-Montes}, Jaime A. and {Archibald}, Anne M. and {B{\'o}di}, Attila and {Bapat}, Shreyas and {Barentsen}, Geert and {Baz{\'a}n}, Juanjo and {Biswas}, Manish and {Boquien}, M{\'e}d{\'e}ric and {Burke}, D.~J. and {Cara}, Daria and {Cara}, Mihai and {Conroy}, Kyle E. and {Conseil}, Simon and {Craig}, Matthew W. and {Cross}, Robert M. and {Cruz}, Kelle L. and {D'Eugenio}, Francesco and {Dencheva}, Nadia and {Devillepoix}, Hadrien A.~R. and {Dietrich}, J{\"o}rg P. and {Eigenbrot}, Arthur Davis and {Erben}, Thomas and {Ferreira}, Leonardo and {Foreman-Mackey}, Daniel and {Fox}, Ryan and {Freij}, Nabil and {Garg}, Suyog and {Geda}, Robel and {Glattly}, Lauren and {Gondhalekar}, Yash and {Gordon}, Karl D. and {Grant}, David and {Greenfield}, Perry and {Groener}, Austen M. and {Guest}, Steve and {Gurovich}, Sebastian and {Handberg}, Rasmus and {Hart}, Akeem and {Hatfield-Dodds}, Zac and {Homeier}, Derek and {Hosseinzadeh}, Griffin and {Jenness}, Tim and {Jones}, Craig K. and {Joseph}, Prajwel and {Kalmbach}, J. Bryce and {Karamehmetoglu}, Emir and {Ka{\l}uszy{\'n}ski}, Miko{\l}aj and {Kelley}, Michael S.~P. and {Kern}, Nicholas and {Kerzendorf}, Wolfgang E. and {Koch}, Eric W. and {Kulumani}, Shankar and {Lee}, Antony and {Ly}, Chun and {Ma}, Zhiyuan and {MacBride}, Conor and {Maljaars}, Jakob M. and {Muna}, Demitri and {Murphy}, N.~A. and {Norman}, Henrik and {O'Steen}, Richard and {Oman}, Kyle A. and {Pacifici}, Camilla and {Pascual}, Sergio and {Pascual-Granado}, J. and {Patil}, Rohit R. and {Perren}, Gabriel I. and {Pickering}, Timothy E. and {Rastogi}, Tanuj and {Roulston}, Benjamin R. and {Ryan}, Daniel F. and {Rykoff}, Eli S. and {Sabater}, Jose and {Sakurikar}, Parikshit and {Salgado}, Jes{\'u}s and {Sanghi}, Aniket and {Saunders}, Nicholas and {Savchenko}, Volodymyr and {Schwardt}, Ludwig and {Seifert-Eckert}, Michael and {Shih}, Albert Y. and {Jain}, Anany Shrey and {Shukla}, Gyanendra and {Sick}, Jonathan and {Simpson}, Chris and {Singanamalla}, Sudheesh and {Singer}, Leo P. and {Singhal}, Jaladh and {Sinha}, Manodeep and {Sip{\H{o}}cz}, Brigitta M. and {Spitler}, Lee R. and {Stansby}, David and {Streicher}, Ole and {{\v{S}}umak}, Jani and {Swinbank}, John D. and {Taranu}, Dan S. and {Tewary}, Nikita and {Tremblay}, Grant R. and {de Val-Borro}, Miguel and {Van Kooten}, Samuel J. and {Vasovi{\'c}}, Zlatan and {Verma}, Shresth and {de Miranda Cardoso}, Jos{\'e} Vin{\'\i}cius and {Williams}, Peter K.~G. and {Wilson}, Tom J. and {Winkel}, Benjamin and {Wood-Vasey}, W.~M. and {Xue}, Rui and {Yoachim}, Peter and {Zhang}, Chen and {Zonca}, Andrea and {Astropy Project Contributors}},
        title = "{The Astropy Project: Sustaining and Growing a Community-oriented Open-source Project and the Latest Major Release (v5.0) of the Core Package}",
      journal = {The Astrophysical Journal},
         year = 2022,
        month = aug,
       volume = {935},
       number = {2},
          eid = {167},
        pages = {167},
          doi = {10.3847/1538-4357/ac7c74},
archivePrefix = {arXiv},
       eprint = {2206.14220},
 primaryClass = {astro-ph.IM},
       adsurl = {https://ui.adsabs.harvard.edu/abs/2022ApJ...935..167A}
}

@article{Kittiwisit:2025,
    author = {Kittiwisit, Piyanat and Murray, Steven G and Garsden, Hugh and Bull, Philip and Wilensky, Michael J and Cain, Christopher and Parsons, Aaron R and Sipple, Jackson and Adams, Tyrone and Aguirre, James E and Baartman, Rushelle and Beardsley, Adam P and Berkhout, Lindsay M and Bernardi, Gianni and Billings, Tashalee S and Bowman, Judd D and Bradley, Richard F and Burba, Jacob and Carey, Steven and Carilli, Chris L and Chen, Kai-Feng and Choudhuri, Samir and Cox, Tyler and DeBoer, David R and Acedo, Eloy de Lera and Dexter, Matt and Dillon, Joshua S and Eksteen, Nico and Ely, John and Ewall-Wice, Aaron and Fagnoni, Nicolas and Furlanetto, Steven R and Gale-Sides, Kingsley and Gehlot, Bharat Kumar and Glendenning, Brian and Gorce, Adelie and Gorthi, Deepthi and Greig, Bradley and Grobbelaar, Jasper and Halday, Ziyaad and Hazelton, Bryna J and Hewitt, Jacqueline N and Hickish, Jack and Jacobs, Daniel C and Josaitis, Alec and Kern, Nicholas S and Kerrigan, Joshua and Kim, Honggeun and Kolopanis, Matthew and Lanman, Adam and Plante, Paul La and Liu, Adrian and Ma, Yin-Zhe and MacMahon, David H E and Malan, Lourence and Malgas, Cresshim and Malgas, Keith and Marero, Bradley and Martinot, Zachary E and McBride, Lisa and Mesinger, Andrei and Molewa, Mathakane and Morales, Miguel F and Mosiane, Tshegofalang and Nunhokee, Chuneeta Devi and Nuwegeld, Hans and Pascua, Robert and Qin, Yuxiang and Rath, Eleanor and Razavi-Ghods, Nima and Robnett, James and Santos, Mario G and Sims, Peter and Singh, Saurabh and Storer, Dara and Swarts, Hilton and Tan, Jianrong and Thyagarajan, Nithyanandan and van Wyngaarden, Pieter and Xu, Zhilei and Zheng, Haoxuan},
    title = {matvis: a matrix-based visibility simulator for fast forward modelling of many-element 21 cm arrays},
    journal = {RAS Techniques and Instruments},
    volume = {4},
    pages = {rzaf001},
    year = {2025},
    month = {01},
    issn = {2752-8200},
    doi = {10.1093/rasti/rzaf001},
    url = {https://doi.org/10.1093/rasti/rzaf001},
    eprint = {https://academic.oup.com/rasti/article-pdf/doi/10.1093/rasti/rzaf001/61413923/rzaf001.pdf},
}

@article{DeBoer:2017,
doi = {10.1088/1538-3873/129/974/045001},
url = {https://dx.doi.org/10.1088/1538-3873/129/974/045001},
year = {2017},
month = {mar},
publisher = {The Astronomical Society of the Pacific},
volume = {129},
number = {974},
pages = {045001},
author = {David R. DeBoer and Aaron R. Parsons and James E. Aguirre and Paul Alexander and Zaki S. Ali and Adam P. Beardsley and Gianni Bernardi and Judd D. Bowman and Richard F. Bradley and Chris L. Carilli and Carina Cheng and Eloy de Lera Acedo and Joshua S. Dillon and Aaron Ewall-Wice and Gcobisa Fadana and Nicolas Fagnoni and Randall Fritz and Steve R. Furlanetto and Brian Glendenning and Bradley Greig and Jasper Grobbelaar and Bryna J. Hazelton and Jacqueline N. Hewitt and Jack Hickish and Daniel C. Jacobs and Austin Julius and MacCalvin Kariseb and Saul A. Kohn and Telalo Lekalake and Adrian Liu and Anita Loots and David MacMahon and Lourence Malan and Cresshim Malgas and Matthys Maree and Zachary Martinot and Nathan Mathison and Eunice Matsetela and Andrei Mesinger and Miguel F. Morales and Abraham R. Neben and Nipanjana Patra and Samantha Pieterse and Jonathan C. Pober and Nima Razavi-Ghods and Jon Ringuette and James Robnett and Kathryn Rosie and Raddwine Sell and Craig Smith and Angelo Syce and Max Tegmark and Nithyanandan Thyagarajan and Peter K. G. Williams and Haoxuan Zheng},
title = {Hydrogen Epoch of Reionization Array (HERA)},
journal = {Publications of the Astronomical Society of the Pacific}
}

@article{Berkhout:2024,
    doi = {10.1088/1538-3873/ad3122},
    url = {https://dx.doi.org/10.1088/1538-3873/ad3122},
    year = {2024},
    month = {apr},
    publisher = {The Astronomical Society of the Pacific},
    volume = {136},
    number = {4},
    pages = {045002},
    author = {Lindsay M. Berkhout and Daniel C. Jacobs and Zuhra Abdurashidova and Tyrone Adams and James E. Aguirre and Paul Alexander and Zaki S. Ali and Rushelle Baartman and Yanga Balfour and Adam P. Beardsley and Gianni Bernardi and Tashalee S. Billings and Judd D. Bowman and Richard F. Bradley and Philip Bull and Jacob Burba and Ruby Byrne and Steven Carey and Chris L. Carilli and Kai-Feng Chen and Carina Cheng and Samir Choudhuri and David R. DeBoer and Eloy de Lera Acedo and Matt Dexter and Joshua S. Dillon and Scott Dynes and Nico Eksteen and John Ely and Aaron Ewall-Wice and Nicolas Fagnoni and Randall Fritz and Steven R. Furlanetto and Kingsley Gale-Sides and Hugh Garsden and Bharat Kumar Gehlot and Abhik Ghosh and Brian Glendenning and Adelie Gorce and Deepthi Gorthi and Bradley Greig and Jasper Grobbelaar and Ziyaad Halday and Bryna J. Hazelton and Jacqueline N. Hewitt and Jack Hickish and Tian Huang and Alec Josaitis and Austin Julius and MacCalvin Kariseb and Nicholas S. Kern and Joshua Kerrigan and Honggeun Kim and Piyanat Kittiwisit and Saul A. Kohn and Matthew Kolopanis and Adam Lanman and Paul La Plante and Adrian Liu and Anita Loots and Yin-Zhe Ma and David Harold Edward MacMahon and Lourence Malan and Cresshim Malgas and Keith Malgas and Bradley Marero and Zachary E. Martinot and Andrei Mesinger and Mathakane Molewa and Miguel F. Morales and Tshegofalang Mosiane and Steven G. Murray and Abraham R. Neben and Bojan Nikolic and Chuneeta Devi Nunhokee and Hans Nuwegeld and Aaron R. Parsons and Robert Pascua and Nipanjana Patra and Samantha Pieterse and Yuxiang Qin and Eleanor Rath and Nima Razavi-Ghods and Daniel Riley and James Robnett and Kathryn Rosie and Mario G. Santos and Peter Sims and Saurabh Singh and Dara Storer and Hilton Swarts and Jianrong Tan and Nithyanandan Thyagarajan and Pieter van Wyngaarden and Peter K. G. Williams and Haoxuan Zheng and Zhilei Xu},
    title = {Hydrogen Epoch of Reionization Array (HERA) Phase II Deployment and Commissioning},
    journal = {Publications of the Astronomical Society of the Pacific}
}

@inproceedings{Newburgh:2016,
author = {L. B. Newburgh and K. Bandura and M. A. Bucher and T.-C. Chang and H. C. Chiang and J.F. Cliche and R. Dav{\'e} and M. Dobbs and C. Clarkson and K. M. Ganga and T. Gogo and A. Gumba and N. Gupta and M. Hilton and B. Johnstone and A. Karastergiou and M. Kunz and D. Lokhorst and R. Maartens and S. Macpherson and M. Mdlalose and K. Moodley and L. Ngwenya and J. M. Parra and J. Peterson and O. Recnik and B. Saliwanchik and M. G. Santos and J. L. Sievers and O. Smirnov and P. Stronkhorst and R. Taylor and K. Vanderlinde and G. Van Vuuren and A. Weltman and A. Witzemann},
title = {{HIRAX: a probe of dark energy and radio transients}},
volume = {9906},
series = {},
booktitle = {Ground-based and Airborne Telescopes VI},
editor = {Helen J. Hall and Roberto Gilmozzi and Heather K. Marshall},
organization = {International Society for Optics and Photonics},
publisher = {SPIE},
pages = {99065X},
year = {2016},
doi = {10.1117/12.2234286},
URL = {https://doi.org/10.1117/12.2234286}
}

@article{Crichton:2022,
author = {Devin Crichton and Moumita Aich and Adam Amara and Kevin Bandura and Bruce A. Bassett and Carlos Bengaly and Pascale Berner and Shruti Bhatporia and Martin Bucher and Tzu-Ching Chang and H. Cynthia Chiang and Jean-Fran{\c{c}}ois Cliche and Carolyn Crichton and Romeel Dave and Dirk I. L. De Villiers and Matt Dobbs and Aaron M. Ewall-Wice and Scott Eyono and Christopher Finlay and Sindhu Gaddam and Ken Ganga and Kevin G. Gayley and Kit Gerodias and Tim B. Gibbon and Austine Gumba and Neeraj Gupta and Maile Harris and Heiko Heilgendorff and Matt Hilton and Adam D. Hincks and Pascal Hitz and Mona Jalilvand and Roufurd P. M. Julie and Zahra Kader and Joseph Kania and Dionysios Karagiannis and Aris Karastergiou and Kabelo Kesebonye and Piyanat Kittiwisit and Jean-Paul Kneib and Kenda Knowles and Emily R. Kuhn and Martin Kunz and Roy Maartens and Vincent MacKay and Stuart MacPherson and Christian Monstein and Kavilan Moodley and V. Mugundhan and Warren Naidoo and Arun Naidu and Laura B. Newburgh and Viraj Nistane and Amanda Di Nitto and Deniz {\"O}l{\c{c}}ek and Xinyu Pan and Sourabh Paul and Jeffrey B. Peterson and Elizabeth Pieters and Carla Pieterse and Aritha Pillay and Anna R. Polish and Liantsoa Randrianjanahary and Alexandre Refregier and Andre Renard and Edwin Retana-Montenegro and Ian H. Rout and Cyndie Russeeawon and Alireza Vafaei Sadr and Benjamin R. B. Saliwanchik and Ajith Sampath and Pranav Sanghavi and Mario G. Santos and Onkabetse Sengate and J. Richard Shaw and Jonathan L. Sievers and Oleg M. Smirnov and Kendrick M. Smith and Ulrich Armel Mbou Sob and Raghunathan Srianand and Pieter Stronkhorst and Dhaneshwar D. Sunder and Simon Tartakovsky and Russ Taylor and Peter Timbie and Emma E. Tolley and Junaid Townsend and Will Tyndall and Cornelius Ungerer and Jacques van Dyk and Gary van Vuuren and Keith Vanderlinde and Thierry Viant and Anthony Walters and Jingying Wang and Amanda Weltman and Patrick Woudt and Dallas Wulf and Anatoly Zavyalov and Zheng Zhang},
title = {{Hydrogen Intensity and Real-Time Analysis Experiment: 256-element array status and overview}},
volume = {8},
journal = {Journal of Astronomical Telescopes, Instruments, and Systems},
number = {1},
publisher = {SPIE},
pages = {011019},
year = {2022},
doi = {10.1117/1.JATIS.8.1.011019},
URL = {https://doi.org/10.1117/1.JATIS.8.1.011019}
}

@article{CHIME:2022,
doi = {10.3847/1538-4365/ac6fd9},
url = {https://dx.doi.org/10.3847/1538-4365/ac6fd9},
year = {2022},
month = {jul},
publisher = {The American Astronomical Society},
volume = {261},
number = {2},
pages = {29},
author = {{CHIME Collaboration} and Mandana Amiri and Kevin Bandura and Anja Boskovic and Tianyue Chen and Jean-François Cliche and Meiling Deng and Nolan Denman and Matt Dobbs and Mateus Fandino and Simon Foreman and Mark Halpern and David Hanna and Alex S. Hill and Gary Hinshaw and Carolin Höfer and Joseph Kania and Peter Klages and T. L. Landecker and Joshua MacEachern and Kiyoshi Masui and Juan Mena-Parra and Nikola Milutinovic and Arash Mirhosseini and Laura Newburgh and Rick Nitsche and Anna Ordog and Ue-Li Pen and Tristan Pinsonneault-Marotte and Ava Polzin and Alex Reda and Andre Renard and J. Richard Shaw and Seth R. Siegel and Saurabh Singh and Rick Smegal and Ian Tretyakov and Kwinten Van Gassen and Keith Vanderlinde and Haochen Wang and Donald V. Wiebe and James S. Willis and Dallas Wulf},
title = {An Overview of CHIME, the Canadian Hydrogen Intensity Mapping Experiment},
journal = {The Astrophysical Journal Supplement Series}
}

@misc{Vanderlinde:2020,
  author       = {Vanderlinde, Keith and
                  Liu, Adrian and
                  Gaensler, Bryan and
                  Bond, Dick and
                  Hinshaw, Gary and
                  Ng, Cherry and
                  Chiang, Cynthia and
                  Stairs, Ingrid and
                  Brown, Jo-Anne and
                  Sievers, Jonathan and
                  Mena, Juan and
                  Smith, Kendrick and
                  Bandura, Kevin and
                  Masui, Kiyoshi and
                  Spekkens, Kristine and
                  Belostotski, Leo and
                  Dobbs, Matt and
                  Turok, Neil and
                  Boyle, Patrick and
                  Rupen, Michael and
                  Landecker, Tom and
                  Pen, Ue-Li and
                  Kaspi, Victoria},
  title        = {{The Canadian Hydrogen Observatory and Radio- 
                   transient Detector (CHORD)}},
  month        = may,
  year         = 2020,
  publisher    = {Zenodo},
  doi          = {10.5281/zenodo.3765414},
  url          = {https://doi.org/10.5281/zenodo.3765414}
}

@ARTICLE{Liu&Shaw:2020,
       author = {{Liu}, Adrian and {Shaw}, J. Richard},
        title = "{Data Analysis for Precision 21 cm Cosmology}",
      journal = {Publications of the Astronomical Society of the Pacific},
         year = 2020,
        month = jun,
       volume = {132},
       number = {1012},
          eid = {062001},
        pages = {062001},
          doi = {10.1088/1538-3873/ab5bfd},
archivePrefix = {arXiv},
       eprint = {1907.08211},
 primaryClass = {astro-ph.IM},
       adsurl = {https://ui.adsabs.harvard.edu/abs/2020PASP..132f2001L}
}
\bibliographystyle{aasjournal}

\appendix

In the appendix, we provide detailed calculations and investigations which are useful for acquiring a deeper understanding of CorrCal but are not necessary for understanding the broader scope of CorrCal.
In~\autoref{sec:SOURCE-COVARIANCE} and~\autoref{sec:DIFFUSE-COVARIANCE}, we explicitly compute the elements of the model covariance matrix (\autoref{eq:covariance-components}) and show how the model covariance can be written as a sum of products of significantly smaller matrices.
This factored representation enables efficient inversion of the covariance, which is an essential ingredient in making CorrCal computationally tractable for large-$N$ arrays.
In~\autoref{sec:ALGORITHM}, we describe how the factored representation of the model covariance is leveraged to efficiently compute the calibration log-likelihood and its gradient.
In~\autoref{sec:MODELING}, we detail an efficient algorithm for computing the diffuse covariance matrix elements within a redundant group.
In~\autoref{sec:PHASE-SLOPE}, we analyze residual phase gradients in the calibration solutions for three qualitatively different observing scenarios and comment on the implications for real applications of CorrCal.
In~\autoref{sec:REDCAL}, we demonstrate how redundant calibration may be recovered as a limiting case of CorrCal.

\section{Point Source Covariance}
\label{sec:SOURCE-COVARIANCE}
Recall from~\autoref{eq:sky-components} that we model point sources as Dirac-delta contributions to the total sky intensity.
Under this assumption, the point source component of the measurement equation (\autoref{eq:model-visibility}) is converted from an integral to a sum, so that
\begin{equation}
    \label{eq:source-model-visibility}
    m_k^{\rm source} = G_k \sum_j s_j A_k(\bvec{r}_j) e^{-i2\pi\nu\mathbfit{b}_k \cdot \bvec{r}_j/c}.
\end{equation}
Because point sources break the statistical symmetry between the real and imaginary parts of the model visibility, we arrange the real and imaginary parts of the model visibility into a length-2 vector and rewrite~\autoref{eq:source-model-visibility} as
\begin{equation}
    \label{eq:source-vis-real-imag-split}
    \begin{pmatrix}
        m_k^{{\rm source},R} \\
        m_k^{{\rm source},I}
    \end{pmatrix} =
    \sum_j s_j
    \begin{pmatrix}
        G_k^R & -G_k^I \\
        G_k^I & G_k^R
    \end{pmatrix}
    \begin{pmatrix}
        f_k(\bvec{r}_j)^R \\
        f_k(\bvec{r}_j)^I
    \end{pmatrix},
\end{equation}
where $f_k(\bvec{r}) \equiv A_k(\bvec{r}) \exp(-i2\pi\nu\mathbfit{b}_k\cdot\bvec{r}/c)$ and the $R, I$ superscripts indicate the real and imaginary parts, respectively.
Using $\mathbfit{m}_k^{\rm source}$, $\mathbf{G}_k$, and $\mathbfit{f}_k(\bvec{r}_j)$ to represent the matrices and vectors in~\autoref{eq:source-vis-real-imag-split}, we may write the model visibility for baseline $\mathbfit{b}_k$ corresponding to point sources as
\begin{equation}
    \mathbfit{m}_k^{\rm source} = \sum_j s_j \mathbf{G}_k \mathbfit{f}_k(\bvec{r}_j) \equiv \mathbf{G}_k \mathbfit{v}_k^{\rm source}.
\end{equation}
Recall from~\autoref{sec:FORMALISM} that we are assuming the source fluxes are mutually independent with zero mean and variance $S_j^2$, so the covariance between the visibility measured by baseline $\mathbfit{b}_k$ and that of baseline $\mathbfit{b}_{k'}$ is
\begin{equation}
    \label{eq:source-covariance-element}
    \mathbf{S}_{kk'} \equiv \Bigl\langle \mathbfit{v}_k^{\rm source} \mathbfit{v}_{k'}^{{\rm source,}T}\Bigr\rangle =\sum_j S_j^2 \mathbfit{f}_k(\bvec{r}_j) \mathbfit{f}_{k'}(\bvec{r}_j)^T,
\end{equation}
where the ensemble average is taken over all realizations of the source fluxes.
Each element of the sum in expression in~\autoref{eq:source-covariance-element}, $S_j^2 \mathbfit{f}_k(\bvec{r}_j) \mathbfit{f}_{k'}(\bvec{r}_j)^T$, has a rather straightforward physical interpretation: the expected power from each source is encoded in $S_j^2$ while the product $\mathbfit{f}_k(\bvec{r}_j) \mathbfit{f}_{k'}(\bvec{r}_j)^T$ encodes how much that power is attenuated by the beam as well as the expected relative phase between the visibilities measured by baselines $\mathbfit{b}_k$ and $\mathbfit{b}_{k'}$.
Furthermore, the sum over sources is well-motivated from a statistical standpoint because the covariance of a sum of random variables is simply the sum of their covariances.

A careful inspection of~\autoref{eq:source-covariance-element} reveals that the point source covariance $\mathbf{S}$ may be factored into a symmetric product of matrices via $\mathbf{S} = \mathbf{\Sigma} \mathbf{\Sigma}^T$.
We refer to $\mathbf{\Sigma}$ as the \emph{source matrix}, because it encodes the interferometric response to each source,
\begin{equation}
    \label{eq:source-matrix}
    \mathbfit{\Sigma}_{kj} = S_j \mathbfit{f}_{k}(\bvec{r}_j).
\end{equation}
The source covariance is evidently \emph{sparse} in the sense that it is fully characterized by the $N_{\rm baseline} \times N_{\rm source}$ terms in the source matrix $\mathbf{\Sigma}$, rather than the full $N_{\rm baseline} \times N_{\rm baseline}$ terms in the full source covariance.
In~\autoref{sec:ALGORITHM}, we demonstrate how this can be leveraged to efficiently invert the model covariance.

\section{Diffuse Covariance}
\label{sec:DIFFUSE-COVARIANCE}
In addition to the covariance between visibilities induced by the presence of point sources on the sky, we must also compute the covariance induced by diffuse emission on the sky.
In this appendix, we provide detailed calculations of how the ``diffuse covariance'' ($\mathbf{D}$ in~\autoref{eq:covariance-components}) may be computed, both in terms of spherical harmonics and in terms of $uv$-modes.
We additionally consider the limiting case of a regular array without any partial redundancy, in which the diffuse covariance becomes block-diagonal.
For the calculations that follow, recall from~\autoref{sec:FORMALISM} that we choose to model the diffuse component of the radio sky as a mean zero, isotropic Gaussian random field.
Consequently, the covariance of the diffuse sky component is fully characterized by a power spectrum that depends only on the angular separation between points on the sky, or equivalently angular multipoles $\ell$ or $uv$-mode magnitudes $|\mathbfit{u}|$.

Recall from~\autoref{sec:FORMALISM} that the statistical symmetry breaking induced in the visibilities by the presence of point sources necessitates that we treat the real and imaginary parts of the visibilities as separate random variables.
Instead of computing the diffuse covariance explicitly using the real and imaginary parts of the visibilities, we instead compute the complex covariance $\mathbf{K} \equiv \bigl\langle \mathbfit{v}^{\rm diffuse}\mathbfit{v}^{\rm diffuse,\dagger} \bigr\rangle$ and pseudo-covariance $\mathbf{\Gamma} \equiv \bigl\langle \mathbfit{v}^{\rm diffuse}\mathbfit{v}^{{\rm diffuse},T} \bigr\rangle$, where $\mathbfit{v}^{\rm diffuse}$ is a vector containing all of the visibilities for the diffuse sky component (excluding the gain terms), and form the covariance between the real and imaginary components through linear combinations of $\mathbf{K}$ and $\mathbf{\Gamma}$.
Explicitly, the diffuse covariance is related to the complex covariance and pseudo-covariance through
\begin{subequations}
    \begin{align}
        2\mathbf{D}^{RR} &= {\rm Re}\bigl( \mathbf{K} + \mathbf{\Gamma}\bigr), \\
        2\mathbf{D}^{II} &= {\rm Re}\bigl( \mathbf{K} - \mathbf{\Gamma}\bigr), \\
        2\mathbf{D}^{RI} &= -{\rm Im}\bigl( \mathbf{K} - \mathbf{\Gamma}\bigr), \\
        2\mathbf{D}^{IR} &= {\rm Im}\bigl( \mathbf{K} + \mathbf{\Gamma}\bigr),
    \end{align}
\end{subequations}
where the $R,I$ superscripts indicate real and imaginary components, respectively (e.g., $D_{kk'}^{RI} \equiv \bigl\langle v_k^R v_{k'}^I \bigr\rangle$).

\subsection{Covariance Modeling in the uv-Plane}
We begin by expressing the diffuse component of the sky emission in the $uv$-plane as the Fourier transform over the image plane of the diffuse sky signal,
\begin{equation}
    \tilde{\mathcal{D}}(u,v) = \int \mathcal{D}(\theta_x, \theta_y) e^{-i2\pi(u\theta_x + v\theta_y)} d\theta_x d\theta_y.
\end{equation}
In the above expression, $\theta_x$ and $\theta_y$ are image plane coordinates in the North-South and East-West directions, respectively, and $\mathcal{D}(\theta_x, \theta_y)$ is the image plane projection of the diffuse emission.
Under the assumption that the diffuse sky signal is a realization of a mean-zero, isotropic Gaussian random field, the $uv$-modes of the signal obey
\begin{equation}
    \bigl\langle \tilde{\mathcal{D}}(\mathbfit{u}) \tilde{\mathcal{D}}(\mathbfit{u}')^*\bigr\rangle = \delta^D(\mathbfit{u} - \mathbfit{u}') P(|\mathbfit{u}|),
\end{equation}
where $\mathbfit{u}^T = (u,v)$ and we refer to $P(|\mathbfit{u}|)$ as the power spectrum of the diffuse sky signal.
Consequently, the two-point correlation of the diffuse sky signal is
\begin{equation}
    \bigl\langle \mathcal{D}(\mathbfit{\theta})\mathcal{D}(\mathbfit{\theta}')\bigr\rangle = \int P(|\mathbfit{u}|) e^{i2\pi\mathbfit{u}\cdot(\mathbfit{\theta}-\mathbfit{\theta}')}dudv,
\end{equation}
where $\mathbfit{\theta}^T = (\theta_x, \theta_y)$.
The above expression will be useful for computing the matrix elements of the diffuse covariance.

To determine the covariance induced in the visibilities, we simply use $\mathcal{D}(\mathbfit{\theta})$ for the sky intensity in the measurement equation (\autoref{eq:model-visibility}), convert the curved sky integral to an integral over the image plane, and compute the complex covariance and pseudo-covariance.
Absorbing the Jacobian for the coordinate transformation into the beam, we get
\begin{align}
    K_{kk'} &= \int P(|\mathbfit{u}|) \Biggl[ \int A_k(\boldsymbol{\theta}) e^{-i2\pi (\mathbfit{u}_k - \mathbfit{u}) \cdot \boldsymbol{\theta}} d\boldsymbol{\theta} \Biggr] \Biggl[\int A_{k'}(\boldsymbol{\theta}') e^{-i2\pi(\mathbfit{u}_{k'} - \mathbfit{u})\cdot\boldsymbol{\theta}'} d\boldsymbol{\theta}' \Biggr]^* d\mathbfit{u}, \\
    \Gamma_{kk'} &= \int P(|\mathbfit{u}|) \Biggl[ \int A_k(\boldsymbol{\theta}) e^{-i2\pi (\mathbfit{u}_k - \mathbfit{u}) \cdot \boldsymbol{\theta}} d\boldsymbol{\theta} \Biggr] \Biggl[\int A_{k'}(\boldsymbol{\theta}') e^{-i2\pi(\mathbfit{u}_{k'} + \mathbfit{u})\cdot\boldsymbol{\theta}'} d\boldsymbol{\theta}' \Biggr] d\mathbfit{u},
\end{align}
where $\mathbfit{u}_k$ is the baseline projected onto the image plane, measured in units of wavelengths.
The terms in brackets are just the phase-shifted Fourier transforms of the power beams, so we may rewrite the above equations as
\begin{align}
    \label{eq:complex-diffuse-covariance}
    K_{kk'} &= \int P(|\mathbfit{u}|) \tilde{A}_k(\mathbfit{u}_k - \mathbfit{u}) \tilde{A}_{k'}(\mathbfit{u}_{k'} - \mathbfit{u})^* d\mathbfit{u}, \\
    \label{eq:complex-diffuse-pseudo-covariance}
    \Gamma_{kk'} &= \int P(|\mathbfit{u}|) \tilde{A}_k(\mathbfit{u}_k - \mathbfit{u}) \tilde{A}_{k'}(\mathbfit{u}_{k'} + \mathbfit{u}) d\mathbfit{u}, 
\end{align}
where $\tilde{A}_k(\mathbfit{u}_k - \mathbfit{u})$ is the \emph{beam kernel} shifted to the $uv$-mode probed by baseline $\mathbfit{b}_k$ (and reflected about the origin).
The above equations mathematically encode a generalized redundant calibration term in the CorrCal model---the diffuse covariance matrix elements evidently only depend on whether the given pair of baselines sample the same $uv$-modes (which is a generalized notion of redundancy) and how much power is contained in fluctuations at the angular scales probed by those baselines.
Because of this, the diffuse covariance effectively primarily provides the redundancy constraints, but in a more flexible way than traditional redundant calibration.

\subsection{Covariance Modeling in Spherical Harmonics}
For completeness, we repeat the calculation from the previous section using the spherical harmonic representation of the sky.
We begin by expanding the diffuse sky signal into spherical harmonics via
\begin{equation}
    \mathcal{D}(\bvec{r}) = \sum_{\ell,m} a_{\ell m} Y_\ell^m(\bvec{r}),
\end{equation}
where $Y_\ell^m(\bvec{r})$ are the spherical harmonic basis functions normalized such that $\int Y_\ell^m(\bvec{r}) Y_{\ell'}^{m'}(\bvec{r}) d\Omega = \delta_{\ell\ell'} \delta_{mm'}$ and the spherical harmonic coefficients $a_{\ell m}$ encode the information about the sky.
For a mean-zero, isotropic Gaussian random field, the spherical harmonic coefficients obey
\begin{equation}
    \bigl\langle a_{\ell m} a_{\ell'm'}^*\bigr\rangle = C_\ell \delta_{\ell\ell'} \delta_{mm'},
\end{equation}
and we refer to $C_\ell$ as the angular power spectrum.
The two-point correlation for the diffuse sky signal is therefore
\begin{equation}
    \Bigl\langle \mathcal{D}(\bvec{r}) \mathcal{D}(\bvec{r}')\Bigr\rangle = \sum_{\ell,m} C_\ell Y_\ell^m(\bvec{r}) Y_\ell^m(\bvec{r}')^*,
\end{equation}
which we can use to obtain the complex covariance and pseudo-covariance via
\begin{align}
    \label{eq:cov-and-pcov-from-harmonics}
    K_{kk'} &= \sum_{\ell,m} C_\ell \Biggl[\int A_k(\bvec{r}) e^{-i2\pi\mathbfit{u}_k \cdot \bvec{r}} Y_\ell^m(\bvec{r}) d\Omega\Biggr] \Biggl[ \int A_{k'}(\bvec{r})^* e^{i2\pi\mathbfit{u}_k \cdot \bvec{r}'} Y_\ell^m(\bvec{r}')^* d\Omega\Biggr], \\
    \Gamma_{kk'} &= \sum_{\ell,m} C_\ell \Biggl[\int A_k(\bvec{r}) e^{-i2\pi\mathbfit{u}_k \cdot \bvec{r}} Y_\ell^m(\bvec{r}) d\Omega\Biggr] \Biggl[ \int A_{k'}(\bvec{r}) e^{-i2\pi\mathbfit{u}_k \cdot \bvec{r}'} Y_\ell^m(\bvec{r}')^* d\Omega\Biggr].
\end{align}
To evaluate the terms in brackets, we expand the beam transfer function $f_k(\bvec{r}) \equiv A_k(\bvec{r})\exp(-i2\pi\mathbfit{u}_k\cdot\bvec{r})$ in spherical harmonics with coefficients $q_{\ell m}^k$ and use the relation $Y_\ell^m(\bvec{r})^* = (-1)^m Y_\ell^{-m}(\bvec{r})$, which ultimately yields
\begin{align}
    K_{kk'} &= \sum_\ell C_\ell \sum_m q_{\ell m}^k q_{\ell m}^{k'*}, \\
    \Gamma_{kk'} &= \sum_\ell C_\ell \sum_m q_{\ell m}^k q_{\ell, -m}^{k'}.
\end{align}
The above representation is the spherical harmonic analog of the final result from the previous section---whether two baselines correlate depends on whether they measure the same $m$-modes, and the amplitude of that correlation is set in part by the amplitude of the power spectrum at the angular scales probed by the baselines.

\subsection{Diffuse Covariance and Array Redundancy}
As shown in the previous section, the covariance between the visibilities measured by two baselines induced by a Gaussian random diffuse signal depends primarily on how much the modes sampled by the two baselines overlap.
Because of the way an interferometer works, the overlap in sampled modes is nonzero only for baselines that are similar to one another.
In physical terms, the voltage measured by an antenna is determined by the coherent sum of the incident electric field taken over the surface of the dish---in essence, the dish itself acts as a continuous interferometric array that is sensitive to modes with $|\mathbfit{u}| \leq D/\lambda$.
Since the measured visibility is the cross-correlation of each antenna's measured voltage, the modes sampled by a baseline are centered at the nominal mode $\mathbfit{u} = \mathbfit{b}/\lambda$ and smeared out by the dish sampling.
Baselines that differ by more than $|\mathbfit{b} - \mathbfit{b}'| \gtrsim \sqrt{2}D$ therefore sample independent modes of the diffuse signal and do not correlate with one another.
In more mathematical terms, the power beam $A_k(\mathbfit{\theta})$ is the product of the voltage beams $\mathbfit{E}_i(\mathbfit{\theta})$,
\begin{equation}
    A_k(\mathbfit{\theta}) = \mathbfit{E}_{k_1}(\mathbfit{\theta}) \cdot \mathbfit{E}_{k_2}(\mathbfit{\theta})^*,
\end{equation}
so the beam kernel is just the convolution of the Fourier transforms of the voltage beams.
The voltage beam for an antenna is just the Fourier transform of the dish illumination pattern, so the beam kernel is the convolution of the individual dish illumination patterns.
Because the dish illumination pattern has a hard edge, so too does the beam kernel, and we arrive at the same conclusion that there is no covariance between the visibilities measured by sufficiently dissimilar baselines.

There are two crucially important conclusions to be made based on the arguments provided in the previous paragraph.
First is that the diffuse covariance serves as a generalized counterpart to redundant calibration, providing constraints based not only on which baselines are redundant but also on pairs of baselines that are partially redundant.
Furthermore, if there are known deviations from redundancy, either in the antenna positions or the beams, then these nonredundancies can be incorporated into the calibration model.
Second is that the diffuse covariance is sparse, since many pairs of baselines in an array will not correlate with one another.
Concretely, when the baselines are sorted into redundant groups, one only needs to keep track of blocks within the diffuse covariance that correspond to pairs of unique baselines whose $uv$ coverage overlap with each other.

In the limiting case where partial redundancies are ignored, the diffuse covariance becomes block-diagonal, enabling a relatively straightforward approach to efficient inversion and compact representation.
In this limit, each \emph{redundant block} $\mathbf{D}_r$ of the diffuse covariance,
\begin{equation}
    \mathbf{D} = {\rm diag}\bigl(\mathbf{D}_1, \cdots, \mathbf{D}_R\bigr),
\end{equation}
where $R$ is the number of redundant groups, may be factored via
\begin{equation}
    \mathbf{D}_r = \mathbf{Q}_r \mathbf{\Lambda}_r^{1/2} \mathbf{\Lambda}_r^{1/2} \mathbf{Q}_r^T,
\end{equation}
where the columns of $\mathbf{Q}_r$ are the eigenvectors for the redundant block and $\mathbf{\Lambda}_r$ is a diagonal matrix containing the corresponding eigenvalues.
This factorization allows us to write the diffuse covariance as a symmetric product of matrices, $\mathbf{D} = \mathbf{\Delta \Delta}^T,$ where the \emph{diffuse matrix} $\mathbf{\Delta}$ is defined as
\begin{equation}
    \mathbf{\Delta} = {\rm diag}\Bigl( \mathbf{Q}_1 \mathbf{\Lambda}_1^{1/2}, \cdots, \mathbf{Q}_R \mathbf{\Lambda}_R^{1/2}\Bigr).
\end{equation}
This representation requires that only a few eigenmodes per redundant group are kept for the calibration model, providing a significant compression of the diffuse covariance.
Details regarding how this representation enables an efficient implementation of the CorrCal algorithm are discussed in depth in~\autoref{sec:ALGORITHM}.

\section{CorrCal Algorithm}
\label{sec:ALGORITHM}
Since the negative log-likelihood we employ to obtain calibration solutions is non-linear in the calibration parameters, we must employ an iterative optimization scheme to obtain the best-fit calibration solution.
We opt to use a conjugate-gradient solver for finding the best-fit calibration solution, so it is imperative that we develop an efficient algorithm for computing the negative log-likelihood and its gradient.
While the addition of the diagonal of the likelihood's Hessian provides an order-of-magnitude reduction in the number of iterations required for convergence, there are numerical challenges with efficiently computing these terms that have yet to be resolved.
Since the current application without the Hessian provides sufficiently accurate and expeditious calibration solutions, we defer the inclusion of the Hessian to future work.
In the remainder of this section, we describe the algorithm we use for computing the negative log-likelihood and its gradient.
We close out the section with a brief discussion of the computational complexity of this implementation of CorrCal.

\subsection{Likelihood Calculation}
\label{sec:LIKELIHOOD}
To perform calibration, we minimize the negative log-likelihood given in~\autoref{eq:negative-log-likelihood}.
Since there is an overall phase degeneracy in the gain solutions, we augment~\autoref{eq:negative-log-likelihood} with an additional Gaussian normalization on the average gain phase so that our negative log-likelihood becomes
\begin{equation}
    \label{eq:nll}
    -\log\mathcal{L} = \log\det \mathbf{C} + \mathbfit{d}^T \mathbf{C}^{-1} \mathbfit{d} + \frac{\big| \sum_a \phi_a \big|^2}{N_{\rm ant}^2 \sigma_{\rm phs}^2},
\end{equation}
where $\phi_a = {\rm arg}(g_a)$ is the phase of antenna $a$, $N_{\rm ant}$ is the number of antennas, and $\sigma_{\rm phs}^2$ is a parameter we may tune to change how strongly this prior is enforced.
Since the number of baselines scales with the square of the number of antennas, the model covariance $\mathbf{C}$ quickly becomes so large that it is computationally prohibitive to work with its dense representation.
Fortunately, it is possible to work with individual matrices that appear in the sparse representation of the covariance in~\autoref{eq:full-sparse-covariance}---we never need to work with the $N_{\rm bl} \times N_{\rm bl}$ dense covariance.

The key to making the log-determinant and matrix inverse calculations in~\autoref{eq:nll} efficient stems from realizing that for a sparse covariance
\begin{equation}
    \label{eq:simple-cov}
    \mathbf{C} = \mathbf{N} + \mathbf{\Delta \Delta}^T + \mathbf{\Sigma \Sigma}^T,
\end{equation}
the inverse covariance also has a sparse representation,
\begin{equation}
    \label{eq:simple-cinv}
    \mathbf{C}^{-1} = \mathbf{N}^{-1} - \bar{\mathbf{\Delta}} \bar{\mathbf{\Delta}}^T - \bar{\mathbf{\Sigma}} \bar{\mathbf{\Sigma}}^T,
\end{equation}
and the log-determinant can be easily accumulated while computing the ``inverse'' diffuse and source matrices $\bar{\mathbf{\Delta}}$ and $\bar{\mathbf{\Sigma}}$.
We stress that the $\bar{\mathbf{\Delta}}$ and $\bar{\mathbf{\Sigma}}$ matrices are \emph{not} the inverses of the diffuse and source matrices.
We simply refer to $\bar{\mathbf{\Delta}}$ and $\bar{\mathbf{\Sigma}}$ as the ``inverse'' diffuse and source matrices because they perform a similar functional role in the sparse representation of the inverse covariance.

The inversion and log-determinant calculation is made computationally efficient by employing repeated applications of the Woodbury identity,
\begin{equation}
    \label{eq:woodbury}
    \bigl(\mathbf{A} + \mathbf{UV}^T\bigr)^{-1} = \mathbf{A}^{-1} - \mathbf{A}^{-1} \mathbf{U}\bigl(\mathbf{I} +  \mathbf{V}^T \mathbf{A}^{-1} \mathbf{U} \bigr)^{-1} \mathbf{V}^T \mathbf{A}^{-1},
\end{equation}
alongside the Matrix Determinant Lemma,
\begin{equation}
    \label{eq:matrix-determinant-lemma}
    \det\bigl(\mathbf{A} + \mathbf{UV}^T\bigr) = \det\bigl(\mathbf{A}\bigr)\det\bigl(\mathbf{I} + \mathbf{V}^T \mathbf{A}^{-1} \mathbf{U}\bigr),
\end{equation}
where in both expressions $\mathbf{I}$ is the identity matrix.
If required, the covariance may be obtained from the inverse covariance by applying the ``inverse'' of the Woodbury identity,
\begin{equation}
    \label{eq:inv-woodbury}
    \bigl(\mathbf{A} - \mathbf{UV}^T\bigr)^{-1} = \mathbf{A}^{-1} + \mathbf{A}^{-1} \mathbf{U}\bigl(\mathbf{I} -  \mathbf{V}^T \mathbf{A}^{-1} \mathbf{U} \bigr)^{-1} \mathbf{V}^T \mathbf{A}^{-1}.
\end{equation}
Applying the Woodbury identity to~\autoref{eq:simple-cov}, we obtain for the ``inverse'' diffuse matrix
\begin{equation}
    \label{eq:inv-diff-mat}
    \bar{\mathbf{\Delta}} = \mathbf{N}^{-1} \mathbf{\Delta} \mathbf{L}_\Delta^{-1T},
\end{equation}
where $\mathbf{L}_\Delta$ is obtained through a Cholesky factorization via
\begin{equation}
    \label{eq:L-delta}
    \mathbf{I} + \mathbf{\Delta}^T \mathbf{N}^{-1} \mathbf{\Delta} = \mathbf{L}_\Delta \mathbf{L}_\Delta^T.
\end{equation}
For the ``inverse'' source matrix $\bar{\mathbf{\Sigma}}$ we find
\begin{equation}
    \label{eq:inv-src-mat}
    \bar{\mathbf{\Sigma}} = \bigl(\mathbf{N}^{-1} - \bar{\mathbf{\Delta}}\bar{\mathbf{\Delta}}^T\bigr) \mathbf{\Sigma} \mathbf{L}_\Sigma^{-1T},
\end{equation}
where $\mathbf{L}_\Sigma$ is defined via
\begin{equation}
    \label{eq:L-sigma}
    \mathbf{I} + \mathbf{\Sigma}^T\bigl(\mathbf{N}^{-1} - \bar{\mathbf{\Delta}} \bar{\mathbf{\Delta}}^T\bigr) \mathbf{\Sigma} = \mathbf{L}_\Sigma \mathbf{L}_\Sigma^T.
\end{equation}
The Cholesky factorizations in~\autoref{eq:L-delta} and~\autoref{eq:L-sigma} contribute negligibly to the total computational cost, since they scale with the cube of the number of eigenmodes and the number of sources, respectively (both of which are small relative to the number of baselines).
Using the Matrix Determinant Lemma, we can easily obtain the log-determinant via
\begin{equation}
    \label{eq:logdet}
    \log\det\mathbf{C} = \log\det\mathbf{N} + 2\log\det\mathbf{L}_\Delta + 2\log\det\mathbf{L}_\Sigma.
\end{equation}
Since the noise matrix $\mathbf{N}$ is independent of the gains, we omit the $\log\det\mathbf{N}$ term when computing the log-determinant.

\subsection{Gradient Calculation}
\label{sec:GRADIENT}
We additionally provide an analytic calculation of the gradient of the negative log-likelihood (\autoref{eq:nll}) with respect to the real and imaginary components of the per-antenna gains,
\begin{equation}
    \label{eq:grad-nll}
    -\partial\log\mathcal{L} = {\rm Tr}\bigl(\mathbf{C}^{-1} \partial \mathbf{C}\bigr) + \mathbfit{d}^T \partial \mathbf{C}^{-1} \mathbfit{d} - \partial\log\mathcal{L}_{\phi},
\end{equation}
where $-\log\mathcal{L}_\phi$ is the phase normalization term.
The gradient of the inverse covariance can be written in terms of the gradient of the covariance via
\begin{equation}
    \label{eq:inv-cov-gradient}
    \partial \mathbf{C}^{-1} = -\mathbf{C}^{-1} \partial \mathbf{C C}^{-1},
\end{equation}
so computing the gradient of the negative log-likelihood ultimately comes down to computing the gradient of the covariance and the gradient of the phase normalization term.
Applying the product rule, the gradient of the covariance consists of four terms,
\begin{align}
    \nonumber
    \partial \mathbf{C} &= \partial \mathbf{G} \mathbf{\Delta \Delta}^T \mathbf{G}^T +  \mathbf{G} \mathbf{\Delta \Delta}^T \partial \mathbf{G}^T \\ &+ \partial \mathbf{G} \mathbf{\Sigma\Sigma}^T \mathbf{G}^T + \mathbf{G} \mathbf{\Sigma \Sigma}^T \partial \mathbf{G}^T.
    \label{eq:full-detail-grad-cov}
\end{align}
Recall from~\autoref{eq:grad-nll} that the gradient of the covariance enters into the negative log-likelihood gradient through a quadratic form, $\mathbfit{d}^T \partial \mathbf{C}^{-1} \mathbfit{d}$, and through a product with a symmetric matrix and a subsequent trace via ${\rm Tr}\bigl(\mathbf{C}^{-1} \partial \mathbf{C}\bigr)$.
Consequently, we may instead use
\begin{equation}
    \partial \mathbf{C} = 2 \partial \mathbf{G} \Bigl(\mathbf{\Delta \Delta}^T + \mathbf{\Sigma \Sigma}^T \Bigr) \mathbf{G}^T
\end{equation}
when computing the gradient of the negative log-likelihood.
Inserting the identity after the gradient of the gain matrix via $\mathbf{I} = \mathbf{G}^{-1} \mathbf{G}$ and simplifying with~\autoref{eq:full-sparse-covariance}, we may rewrite the above expression as
\begin{equation}
    \label{eq:simplified-covariance-gradient}
    \partial \mathbf{C} = 2 \partial\mathbf{G G}^{-1}\bigl(\mathbf{C} - \mathbf{N}\bigr).
\end{equation}

With the simplified expression for the covariance gradient, the gradient of the $\chi^2$ term, $\partial\chi^2 = \mathbfit{d}^T \partial \mathbf{C}^{-1} \mathbfit{d}$, can be written as
\begin{equation}
    \mathbfit{d}^T \partial \mathbf{C}^{-1} \mathbfit{d} = -2\mathbfit{p}^T \partial \mathbf{G} \mathbfit{q},
\end{equation}
where $\mathbfit{p} \equiv \mathbf{C}^{-1} \mathbfit{d}$ and $\mathbfit{q} \equiv \mathbf{G}^{-1} \bigl(\mathbf{C} - \mathbf{N}\bigr) \mathbfit{p}$.
Expanding the model covariance with~\autoref{eq:full-sparse-covariance} and simplifying, the vector $\mathbfit{q}$ can equivalently be written as $\mathbfit{q} = \bigl( \mathbf{\Delta \Delta}^T + \mathbf{\Sigma \Sigma}^T \bigr) \mathbf{G}^T \mathbfit{p}$.
Since the gain matrix is block-diagonal with $2\times2$ blocks
\begin{equation}
    \mathbf{G}_k = \begin{pmatrix}
        G_k^R & -G_k^I \\
        G_k^I & G_k^R
    \end{pmatrix},
\end{equation}
the $\chi^2$ gradient is just a sum over baselines,
\begin{equation}
    \mathbfit{d}^T \partial \mathbf{C}^{-1} \mathbfit{d} = -2\sum_k \mathbfit{p}_k^T \partial \mathbf{G}_k \mathbfit{q}_k,
\end{equation}
where $\mathbfit{p}_k^T = \Bigl(p_k^R,  p_k^I\Bigr)$ and $\mathbfit{q}_k^T = \Bigl(q_k^R,  q_k^I\Bigr)$.
Evaluating the quadratic form allows us to rewrite the $\chi^2$ gradient as
\begin{equation}
    \label{eq:chisq-gradient}
    \mathbfit{d}^T \partial {\bf C}^{-1} \mathbfit{d} = -2 \sum_k s_k \partial G_k^R + t_k \partial G_k^I,
\end{equation}
where $s_k = p_k^R q_k^R + p_k^I q_k^I$ and $t_k = p_k^I q_k^R - p_k^R q_k^I$.
The gradient of the $\chi^2$ term may therefore be reduced to a single sum over baselines.

The trace term in~\autoref{eq:grad-nll} may also be written as a sum over baselines.
To see this, we first insert the simplified expression for the gradient of the covariance from~\autoref{eq:simplified-covariance-gradient} into the trace term to obtain
\begin{equation}
    {\rm Tr}\bigl({\bf C}^{-1} \partial{\bf C}\bigr) = 2 {\rm Tr}\bigl( {\bf C}^{-1} \partial {\bf G G}^{-1} ({\bf C} - {\bf N})\bigr).
\end{equation}
Applying the cyclic property of the trace and distributing the inverse covariance, we may rewrite the previous expression as
\begin{equation}
    {\rm Tr}\bigl({\bf C}^{-1} \partial {\bf C}\bigr) = 2 {\rm Tr}\Bigl( \partial {\bf GG}^{-1} \bigl({\bf I} - {\bf NC}^{-1}\bigr)\Bigr).
\end{equation}
Inserting the sparse representation of the inverse covariance (\autoref{eq:simple-cinv}), the trace term then becomes
\begin{equation}
    {\rm Tr}\bigl( {\bf C}^{-1} \partial {\bf C}\bigr) = 2{\rm Tr}\Bigl( \partial {\bf GG}^{-1} {\bf N} \bigl(\bar{\bf \Delta} \bar{\bf \Delta}^T + \bar{\bf \Sigma} \bar{\bf \Sigma}^T\bigr)\Bigr).
\end{equation}
Since the gain matrix is block-diagonal and the noise matrix is diagonal, the trace is just a sum of traces of products of $2\times2$ matrices,
\begin{equation}
    \label{eq:grad-logdet-almost-final}
    {\rm Tr}\bigl({\bf C}^{-1} \partial {\bf C}\bigr) = 2\sum_k {\rm Tr}\Bigl( \partial {\bf G}_k {\bf G}_k^{-1} {\bf N}_k \bar{\bf P}_k\Bigr),
\end{equation}
where the matrix $\bar{\bf P}_k$ is defined as
\begin{equation}
    \bar{\bf P}_k = \sum_\lambda \begin{pmatrix}
        \bigl(\bar{\Delta}_{k\lambda}^R)^2 & \bar{\Delta}_{k\lambda}^R \bar{\Delta}_{k\lambda}^I \\
        \bar{\Delta}_{k\lambda}^R \bar{\Delta}_{k\lambda}^I & \bigl(\bar{\Delta}_{k\lambda}^I\bigr)^2
    \end{pmatrix}
    + \sum_j \begin{pmatrix}
        \bigl( \bar{\Sigma}_{kj}^R\bigr)^2 & \bar{\Sigma}_{kj}^R \bar{\Sigma}_{kj}^I \\
        \bar{\Sigma}_{kj}^R \bar{\Sigma}_{kj}^I & \bigl(\bar{\Sigma}_{kj}^I\bigr)^2
    \end{pmatrix}.
\end{equation}
Note that $\bar{\bf P}_k$ has the generic structure
\begin{equation}
    \bar{\bf P}_k = \begin{pmatrix} a_k & b_k \\ b_k & c_k\end{pmatrix},
\end{equation}
whereas the matrix product $\partial {\bf GG}^{-1}$ has the generic structure
\begin{equation}
    \partial {\bf G}_k {\bf G}_k^{-1} = \begin{pmatrix} d_k & e_k \\ -e_k & d_k\end{pmatrix}.
\end{equation}
Since ${\bf N}_k = \sigma_k^2 {\bf I}$, each trace in~\autoref{eq:grad-logdet-almost-final} generically looks like
\begin{equation}
    {\rm Tr}\Bigl( \partial{\bf G}_k {\bf G}_k^{-1} {\bf N}_k \bar{\bf P}_k\Bigr) = \sigma_k^2 d_k(a_k+c_k).
\end{equation}
Defining the ``inverse power'' $\bar{P}_k$ as
\begin{equation}
    \label{eq:inverse-power}
    \bar{P}_k = \sum_\lambda \bigl(\bar{\Delta}_{k\lambda}^R\bigr)^2 + \bigl(\bar{\Delta}_{k\lambda}^I\bigr)^2 + \sum_j \bigl(\bar{\Sigma}_{kj}^R\bigr)^2 + \bigl(\bar{\Sigma}_{kj}^I\bigr)^2,
\end{equation}
and noting that ${\bf G}_k^{-1} = |G_k|^{-2} {\bf G}_k^T$, with $|G_k|^2 = \bigl(G_k^R\bigr)^2 + \bigl(G_k^I\bigr)^2$, the trace may be written as a sum over baselines
\begin{equation}
    \label{eq:logdet-gradient}
    {\rm Tr}\bigl( {\bf C}^{-1} \partial {\bf C}\bigr) = 2\sum_k \frac{ \sigma_k^2 \bar{P}_k}{|G_k|^2} \Bigl( G_k^R \partial G_k^R + G_k^I \partial G_k^I\Bigr).
\end{equation}

In order to evaluate the $\chi^2$ gradient (\autoref{eq:chisq-gradient}) and the log-determinant gradient (\autoref{eq:logdet-gradient}), we must compute the gradients with respect to the gain matrix elements $G_k^R$ and $G_k^I$.
Recall that the complex gain matrix elements take the form
\begin{equation}
    G_k = g_{k_1} g_{k_2}^*,
\end{equation}
where $k_1, k_2$ denote the antennas that form baseline $k$.
The gain matrix elements $G_k^R$ and $G_k^I$ are therefore
\begin{subequations}
    \begin{align}
        G_k^R &= g_{k_1}^R g_{k_2}^R + g_{k_1}^I g_{k_2}^I, \\
        G_k^I &= g_{k_1}^I g_{k_2}^R - g_{k_1}^R g_{k_2}^I.
    \end{align}
\end{subequations}
Since the calibration parameters are the real and imaginary parts of the gains, we have four terms to compute, which are
\begin{subequations}
    \begin{align}
    \label{eq:gain-mat-grad-start}
        \frac{\partial G_k^R}{\partial g_a^R} &= g_{k_1}^R \delta_{ak_2} + g_{k_2}^R \delta_{ak_1}, \\
        \frac{\partial G_k^R}{\partial g_a^I} &= g_{k_1}^I \delta_{ak_2} + g_{k_2}^I \delta_{ak_1}, \\
        \frac{\partial G_k^I}{\partial g_a^R} &= g_{k_1}^I \delta_{ak_2} - g_{k_2}^I \delta_{ak_1}, \\
        \frac{\partial G_k^I}{\partial g_a^I} &= g_{k_2}^R \delta_{ak_1} - g_{k_1}^R \delta_{ak_2}.
    \label{eq:gain-mat-grad-end}
    \end{align}
\end{subequations}
Since a given baseline only contributes to the gradient for two antennas, we may efficiently accumulate the gradient by looping over baselines and using~\autoref{eq:gain-mat-grad-start}--\autoref{eq:gain-mat-grad-end} to determine which components of the gradient each baseline contributes to.

In addition to the $\chi^2$ and log-determinant gradients, we must also compute the gradient of the phase normalization term,
\begin{equation}
    -\log\mathcal{L}_\phi = \frac{\Big|\sum_a \phi_a \Big|^2}{N_{\rm ant}^2 \sigma_{\rm phs}^2}.
\end{equation}
Since the gain phases are related to the real and imaginary components of the gains through $\phi_a = \tan^{-1}(g_a^I/g_a^R)$, the gradient of the phase normalization term is
\begin{equation}
    -\partial\log\mathcal{L}_\phi = 2 \frac{\sum_a \phi_a}{N_{\rm ant}^2 \sigma_{\rm phs}^2} \sum_a\frac{1}{1 + \bigl(g_a^I/g_a^R\bigr)^2} \partial\bigl(g_a^I/g_a^R\bigr).
\end{equation}
Writing $|g_a|^2 = \bigl(g_a^R\bigr)^2 + \bigl(g_a^I\bigr)^2$, we may simplify this to
\begin{equation}
    -\partial\log\mathcal{L}_\phi = 2\frac{\sum_a \phi_a}{N_{\rm ant}^2 \sigma_{\rm phs}^2} \sum_a \frac{\bigl(g_a^R\bigr)^2}{|g_a|^2} \partial\bigl(g_a^I/g_a^R\bigr).
\end{equation}
Noting that $\sin\phi_a = g_a^I / |g_a|$ and $\cos\phi_a = g_a^R / |g_a|$, the phase normalization gradient may therefore be written as
\begin{subequations}
    \begin{align}
        -\frac{\partial\log\mathcal{L}_\phi}{\partial g_b^R} &= -2 \frac{\sum_a \phi_a}{N_{\rm ant}^2 \sigma_{\rm phs}^2} \frac{\sin\phi_b}{|g_b|}, \\
        -\frac{\partial\log\mathcal{L}_\phi}{\partial g_b^I} &= 2\frac{\sum_a \phi_a}{N_{\rm ant}^2 \sigma_{\rm phs}^2} \frac{\cos\phi_b}{|g_b|}.
    \end{align}
\end{subequations}
This completes the set of equations needed to analytically compute the gradient of the negative log-likelihood (\autoref{eq:grad-nll}).

\section{Efficient Diffuse Covariance Computation}
\label{sec:MODELING}
Recall from~\autoref{sec:DIFFUSE-COVARIANCE} that the diffuse covariance element $D_{kk'}$ is computed by integrating the product of the beam kernels weighted against the diffuse sky power spectrum.
This fairly expensive calculation, which must be evaluated for every pair of baselines that contribute to the diffuse covariance, can be significantly sped up for pairs of baselines within a given redundant group.
The motivating idea behind the calculation detailed below is that in many scenarios the beam kernels have compact support on a small region of the $uv$-plane, over which the power spectrum should only slowly vary, and this can be leveraged to greatly reduce the computational demands of building the calibration model. 

To simplify the calculation of the overlap integral, we first rewrite the diffuse sky power spectrum $P(|\mathbfit{u}|)$ in terms of band powers $p_\alpha$ via
\begin{equation}
    P(|\mathbfit{u}|) = \sum_\alpha p_\alpha \mathcal{H}_\alpha(|\mathbfit{u}|),
\end{equation}
where
\begin{equation}
    \mathcal{H}_\alpha(|\mathbfit{u}|) =
    \begin{cases}
        1, \qquad \Big||\mathbfit{u}| - |\mathbfit{u}_\alpha|\Big| \leq \epsilon \\
        0, \qquad {\rm else}
    \end{cases},
\end{equation}
for some $\epsilon > 0$.
Expressing the power spectrum in this way may be thought of as binning the power spectrum into $|\mathbfit{u}|$-bins of equal width $2\epsilon$.
For baselines $k$ and $k'$ in redundant group $r$, expressing the power spectrum in terms of band powers enables us to rewrite the overlap integral from~\autoref{eq:complex-diffuse-covariance} as
\begin{equation}
    I_{kk'} = p_{\alpha_r} \int_{|\mathbfit{u}_r|-\epsilon}^{|\mathbfit{u}_r|+\epsilon} \tilde{A}_k(\mathbfit{u}_k-\mathbfit{u}) \tilde{A}_{k'}(\mathbfit{u}_{k'}-\mathbfit{u})^* d\mathbfit{u},
\end{equation}
where $\alpha_r$ is the bin containing the redundant baseline, $\mathbfit{u}_r \equiv N_{k\in r}^{-1}\sum_{k \in r} \mathbfit{u}_k$ is the representative baseline for the redundant group, and $N_{k\in r}$ is the number of baselines in the redundant group.
Since the beam kernel $\tilde{A}_k(\mathbfit{u}_k-\mathbfit{u})$ is nonzero only within some small region around $\mathbfit{u}_k$, the overlap integral is effectively taken over a disk of radius $\epsilon$ centered on $\mathbfit{u}_r$, so we may perform a change of variable to rewrite the integral as
\begin{equation}
    \label{eq:diffuse-cov-element-simplified}
    I_{kk'} = p_{\alpha_r} \int_{|\mathbfit{u}|\leq\epsilon} \tilde{A}_k(\mathbfit{u} - \delta\mathbfit{u}_{kr}) \tilde{A}_{k'}(\mathbfit{u} - \delta\mathbfit{u}_{k'r})^* d\mathbfit{u},
\end{equation}
where $\delta\mathbfit{u}_{kr} \equiv \mathbfit{u}_r - \mathbfit{u}_k$.
Because the beam kernels have compact support in the $uv$-plane, we may approximate the above integral as being taken over the entire $uv$-plane and rewrite the integral as a Fourier transform via
\begin{equation}
    I_{kk'} \approx p_{\alpha_r} \int A_k(\mathbfit{\theta}) A_{k'}(\mathbfit{\theta})^* e^{-i2\pi\delta\mathbfit{u}_{kk'}\cdot\mathbfit{\theta}}d\mathbfit{\theta},
\end{equation}
where $\delta\mathbfit{u}_{kk'} = \mathbfit{u}_k - \mathbfit{u}_{k'}$.
Returning to a curved sky integral, we get
\begin{equation}
    I_{kk'} \approx p_{\alpha_r} \int A_k(\bvec{r}) A_{k'}(\bvec{r})^* e^{-i2\pi\delta\mathbfit{u}_{kk'}\cdot\bvec{r}} d\Omega,
\end{equation}
which we may simplify by expanding the beam product in spherical harmonics via
\begin{equation}
    A_k(\bvec{r}) A_{k'}(\bvec{r})^* = \sum_{\ell,m} B_{\ell m}^{kk'} Y_\ell^m(\bvec{r}),
\end{equation}
and using the spherical harmonic expansion of a plane wave,
\begin{equation}
    e^{-i\mathbfit{s}\cdot\mathbfit{r}} = 4\pi \sum_{\ell, m} i^\ell j_\ell(sr) Y_\ell^m(\bvec{s}) Y_\ell^m(\bvec{r})^*,
\end{equation}
where $s = |\mathbfit{s}|$, $\bvec{s} = \mathbfit{s}/s$, and $j_\ell$ is the spherical Bessel function of the first kind of order $\ell$.
The spherical harmonic expansions allow us to rewrite the integral as a sum over multipoles,
\begin{equation}
    \label{eq:diffuse-covariance-harmonic-sum}
    I_{kk'} \approx 4\pi p_{\alpha_r} \sum_{\ell,m} B_{\ell m}^{kk'} i^\ell j_\ell\bigl(2\pi|\delta\mathbfit{u}_{kk'}|\bigr) Y_\ell^m\biggl(\frac{\delta\mathbfit{u}_{kk'}}{|\delta\mathbfit{u}_{kk'}|}\biggr).
\end{equation}
Since the Bessel functions tend to decay rapidly with $\ell$ for typical values of $|\delta\mathbfit{u}_{kk'}|$, it is usually sufficient to truncate this sum at $\ell \sim 10$ and obtain accurate estimates of the diffuse covariance elements.
Furthermore, if beam nonredundancies are not known \emph{a priori}, then the beam harmonics $B_{\ell m}^{kk'}$ only need to be computed once per redundant group, rather than separately for each pair of baselines, which provides an additional reduction in computation time.

\section{Phase Gradients}
\label{sec:PHASE-SLOPE}
\begin{figure*}
    \includegraphics[width=\textwidth]{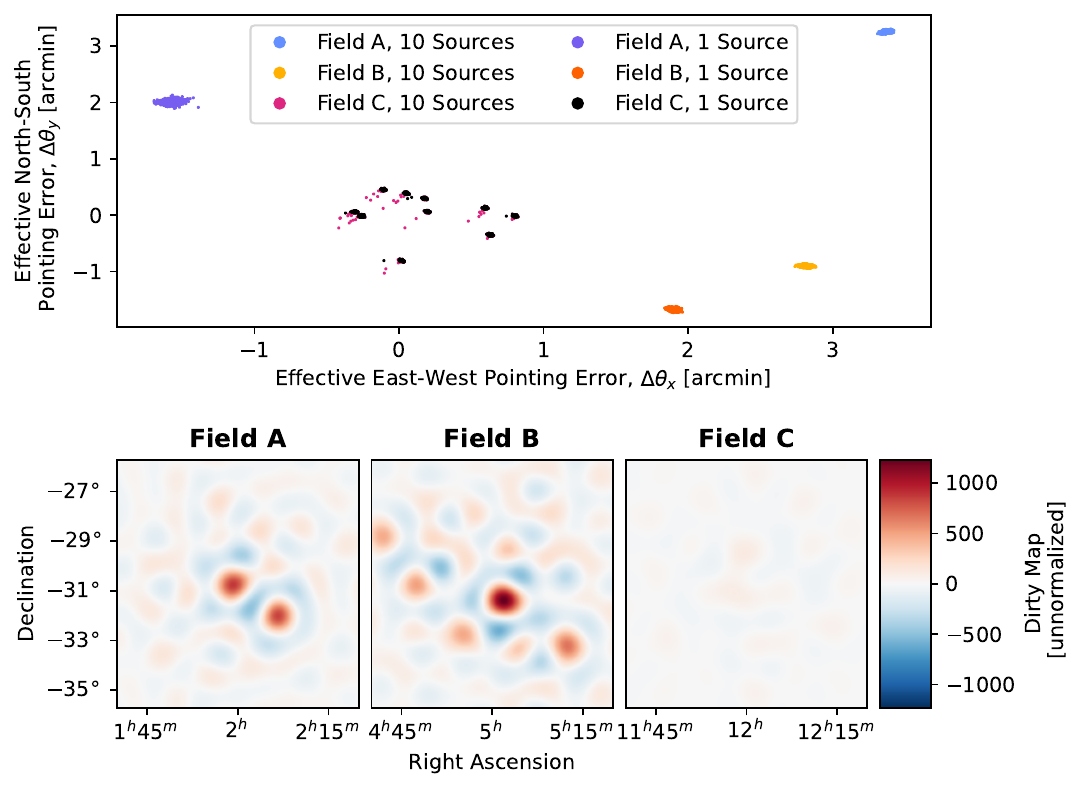}
    \caption{
        Demonstration that CorrCal constrains the phase gradient when there are bright point sources overhead.
        The bottom panels shows the sky, as seen by the array (i.e., imaged by the array, but unnormalized), for three different fields used in the test.
        Field A (bottom left) and Field B (bottom middle) both contain bright point sources that provide a tight constraint on the gain phases.
        Field C (bottom right), however, does not contain any bright sources overhead.
        The top panel shows the phase gradient in the calibration solutions, converted to an effective pointing error, obtained with one calibration source and with ten calibration sources for each field across 100 independent noise realizations and 10 different initial guesses of the gains.
        For Field A and Field B, the phase gradients converge to values that depend on the number of calibration sources used.
        For Field C, the phase gradient is unconstrained, evidenced by the presence of multiple clusters of solutions, each of which correspond to a different initial guess of the gains.
    }
    \label{fig:phase-gradient-summary}
\end{figure*}

Practical applications of CorrCal will typically only use $\mathcal{O}(10)$ calibration sources when constructing the source matrix (\autoref{eq:source-matrix}).
There are two main motivations for this choice, but what makes this a compelling choice is how these two factors compound with one another.
On the one hand, the computational complexity of the CorrCal algorithm scales quadratically with the number of sources; on the other hand, the quality of a calibration solution is only marginally improved with the addition of more sources beyond those that dominate the measured visibilities.
Taken together, modeling more than just the few brightest sources has the effect of making the calibration significantly more expensive while not significantly improving the calibration.
It is therefore imperative that we understand the consequences of calibrating with an incomplete catalog of point sources when constructing the source matrix.
While the expected effect of calibrating with an incomplete source catalog is not analytically tractable, we glean some insight into this issue through simulations, the results of which are summarized in~\autoref{fig:phase-gradient-summary} and discussed later on in this section.

The main effect that an incomplete calibration model has on the gain solutions is the introduction of a spatial phase gradient.
More precisely, the best-fit gains $\hat{g}_a$ will differ from the true gains $g_a^{\rm true}$ through a phase factor that depends on the position of the antenna in the array $\mathbfit{x}_a$ via
\begin{equation}
    \hat{g}_a = g_a^{\rm true} e^{-i\nabla\Phi \cdot \mathbfit{x}_a},
\end{equation}
where we refer to $\nabla\Phi^T = (\partial_x\Phi, \partial_y\Phi)$ as the \emph{phase gradient} in the calibration solutions.
In terms of the calibrated visibilities, this manifests as an effective pointing error, since
\begin{equation}
    \frac{g_i^{\rm true} (g_j^{\rm true})^*}{\hat{g}_i \hat{g}_j^*}V_{ij} = \int A(\boldsymbol{\theta}) I(\boldsymbol{\theta}) e^{-i(2\pi\nu\boldsymbol{\theta}/c - \nabla\Phi)\cdot\mathbfit{b}_{ij}}d\boldsymbol{\theta}.
\end{equation}
Calibrating the data using gains that contain a phase gradient therefore shifts the inferred sky by an amount
\begin{equation}
    \Delta\boldsymbol{\theta} = \frac{\lambda}{2\pi}\nabla\Phi,
\end{equation}
where $\lambda = c/\nu$ is the observed wavelength.
In redundant calibration, the phase gradient is a genuine degeneracy in the calibration solutions~\citep{Liu:2010,Dillon:2018} that must be resolved with an additional absolute calibration step.
In CorrCal, the inclusion of bright point sources breaks this degeneracy; however, this does not mean that the CorrCal solutions are free of a phase gradient, since the missing flux from the source model is absorbed into the calibration solutions as a preferred phase gradient.

\autoref{fig:phase-gradient-summary} summarizes the results of a set of tests performed to demonstrate the behavior of the phase gradient in CorrCal solutions.
These tests were designed to demonstrate that the phase gradient is a function of both the flux on the sky and the sources used for calibration, and also show that the phase gradient becomes unconstrained when there is not a bright source overhead to serve as a reliable calibrator.
With this in mind, we simulated visibilities for three different fields, which we visualize as maps made from the data in the bottom panels of~\autoref{fig:phase-gradient-summary}.
In two of the fields, there are bright sources very close to zenith, while for the third field there are no bright sources overhead.
For each of these three fields, we generated 100 independent samples of radiometer noise with a similar signal-to-noise ratio as we used for the tests in~\autoref{sec:PTSRC-TESTS}, providing us with 300 sets of visibilities to calibrate.
We used a single set of per-antenna gains across all the calibration runs as the true gains that were applied to the data, and randomly generated 10 different sets of initial guesses of the gains using the procedure summarized in~\autoref{table:sim-gain-parameters}.
For each realization of noise, initial guess, and field, we ran calibration with two different covariance models: one covariance model used the ten brightest observed sources, while the other used the brightest observed source.
After obtaining the calibration solutions, we performed a linear least-squares fit to determine the phase gradients by comparing the recovered gain phases against the true gain phases.

The effective pointing errors associated with the recovered phase gradients are plotted in the top panel of~\autoref{fig:phase-gradient-summary}.
For the two fields that contain bright sources, the phase gradients cluster around particular values of $\Delta\boldsymbol{\theta}$ that differ between fields and between covariance models regardless of the initial guess for the gains.
\autoref{fig:phase-gradient-summary} therefore clearly demonstrates that the inclusion of bright point sources induces a preferred phase gradient in the gain solutions.
Importantly, however, the induced pointing error is several times smaller than the synthesized beam for the array---a visual inspection of the bottom row of~\autoref{fig:phase-gradient-summary} suggests the synthesized beam is roughly a degree across, while the phase gradient clusters in the top panel correspond to arcminute-scale pointing errors.
For the field without bright sources, the phase gradients are scattered in 10 groups that correspond to the 10 different initial guesses for the gains---for this field, the calibration solutions settle into some local minimum that is determined by the initial guess.
Evidently, in the absence of bright calibrator sources, the phase gradient is unconstrained by CorrCal.
This should be expected, since in the absence of bright sources CorrCal effectively operates as a generalized version of redundant calibration.
In the version of CorrCal investigated in this paper, where different frequency channels and different integrations are calibrated independently,~\autoref{fig:phase-gradient-summary} indicates that a modest phase gradient calibration may be required for precision cosmology applications.
In a more generalized application, however, it seems plausible that leveraging known covariances in time and frequency to perform a joint fit across multiple times or frequencies could provide enough additional constraining power to produce calibration solutions without a phase gradient.

\vspace{2\baselineskip}
\section{Redundant Calibration Limit}
\label{sec:REDCAL}
In this appendix, we show how an appropriate choice of parameters for the model covariance results in the CorrCal chi-squared, $\chi^2_C$, converging to the redundant calibration chi-squared, $\chi^2_R$, evaluated at the best-fit visibility solutions.
The redundant calibration chi-squared may be written as
\begin{equation}
    \label{eq:redcal-chisq}
    \chi^2_R = \sum_r \sum_{k \in r} \frac{\big|d_k - G_k V_r \big|^2}{\sigma_k^2},
\end{equation}
where $V_r$ is the model visibility for redundant group $r$, $\sum_{k \in r}$ indicates a sum over all baselines within redundant group $r$, $d_k$ is the visibility data for baseline $\mathbf{b}_k$, $\sigma_k^2$ is the noise variance, and $G_k \equiv g_{k_1} g_{k_2}^*$ is the product of the complex per-antenna gains.
The gradient of $\chi^2_R$ with respect to the model visibilities $V_r$ may be written as
\begin{equation}
    \frac{\partial \chi^2_R}{\partial V_r} = -\sum_{k \in r} \frac{(G_k d_k^* - |G_k|^2 V_r^*)}{\sigma_k^2},
\end{equation}
so the model visibilities at the minimum $\chi^2_R$ are related to the gains via
\begin{equation}
    \hat{V}_r = \frac{\sum_{k \in r} G_k^* d_k / \sigma_k^2}{\sum_{k \in r} |G_k|^2 / \sigma_k^2}.
\end{equation}
If we define the parameters $\gamma_r \equiv \sum_{k \in r} G_k^* d_k / \sigma_k^2$ and $\beta_r \equiv \sum_{k \in r} |G_k|^2 / \sigma_k^2$, then the best-fit model visibilities are just $\hat{V}_r = \gamma_r / \beta_r$.
Inserting the best-fit model visibilities into~\autoref{eq:redcal-chisq}, we get
\begin{align}
    \chi^2_R \bigg|_{V_r=\hat{V}_r} = \sum_r \sum_{k \in r}\frac{|d_k - G_k \gamma_r / \beta_r|^2}{\sigma_k^2}, \\
\end{align}
which may be simplified as
\begin{equation}
    \label{eq:best-fit-redcal-chisq}
    \chi^2_R \bigg|_{V_r = \hat{V}_r} = \sum_k \frac{|d_k|^2}{\sigma_k^2} - \sum_r \frac{|\gamma_r|^2}{\beta_r}.
\end{equation}
This is exactly equal to the CorrCal chi-squared in a particular limiting case, as we will show below.

In the case that no point sources are included in the model covariance, the data may be treated as circularly symmetric Gaussian random variables, and we may therefore use the complex-valued data and complex-valued model covariance so that
\begin{equation}
    \chi^2_C = \mathbfit{d}^\dagger \mathbf{C}^{-1} \mathbfit{d},
\end{equation}
where $\mathbfit{d}^T = (d_1, \cdots, d_N)$, and $\mathbf{C} = \mathbf{N} + \mathbf{G\Delta\Delta}^\dagger \mathbf{G}^\dagger$.
Note that in this representation, the noise matrix $\mathbf{N}$ and gain matrix $\mathbf{G}$ are both diagonal.
For the diffuse matrix, we use a block-diagonal parametrization with
\begin{equation}
    \mathbf{\Delta} = {\rm diag}\Bigl( \cdots, V_r \mathbf{1}, \cdots\Bigr),
\end{equation}
where $\mathbf{1}$ is a vector of ones and $V_r$ is the model visibility for redundant group $r$.
In order to compute $\chi^2_C$, we will need to manually compute $\mathbf{C}^{-1}$, which we can do by applying the Woodbury identity so that
\begin{equation}
    \mathbf{C}^{-1} = \mathbf{N}^{-1} - \mathbf{N}^{-1} \mathbf{G\Delta} \bigl( \mathbf{I} + \mathbf{\Delta}^\dagger \mathbf{G}^\dagger \mathbf{N}^{-1} \mathbf{G\Delta}\bigr)^{-1} \mathbf{\Delta}^\dagger \mathbf{G}^\dagger \mathbf{N}^{-1},
\end{equation}
where $\mathbf{I}$ is the identity matrix.
Since $\mathbf{\Delta}$ is block-diagonal, and both $\mathbf{G}$ and $\mathbf{N}$ are diagonal, the small inverse in the above expression may be written as
\begin{equation}
    \Bigl[\bigl(\mathbf{I} + \mathbf{\Delta}^\dagger \mathbf{G}^\dagger \mathbf{N}^{-1} \mathbf{G \Delta}\bigr)^{-1}\Bigr]_{rr'} = \frac{\delta_{rr'}}{1 + |V_r|^2 \sum_{k \in r} |G_k|^2/\sigma_k^2},
\end{equation}
which may be simplified as
\begin{equation}
    \Bigl[\bigl(\mathbf{I} + \mathbf{\Delta}^\dagger \mathbf{G}^\dagger \mathbf{N}^{-1} \mathbf{G\Delta}\bigr)^{-1}\Bigr]_{rr'} = \delta_{rr'} \bigl( 1 + |V_r|^2 \beta_r \bigr)^{-1}.
\end{equation}
The final ingredient to compute the CorrCal $\chi^2$ in this limit is the vector $\mathbf{\Delta}^\dagger \mathbf{G}^\dagger \mathbf{N}^{-1} \mathbfit{d}$.
Since $\mathbf{G}$ and $\mathbf{N}$ are diagonal, $\mathbf{G}^\dagger \mathbf{N}^{-1} \mathbfit{d}$ just scales each component of $\mathbfit{d}$, and contracting the result with $\mathbf{\Delta}^\dagger$ results in sums over redundant groups so that
\begin{equation}
    \Bigl(\mathbf{\Delta}^\dagger \mathbf{G}^\dagger \mathbf{N}^{-1} \mathbfit{d}\Bigr)_r = V_r^* \sum_{k \in r} \frac{G_k^* d_k}{\sigma_k^2},
\end{equation}
which we may simplify as
\begin{equation}
    \Bigl(\mathbf{\Delta}^\dagger \mathbf{G}^\dagger \mathbf{N}^{-1} \mathbfit{d}\Bigr)_r = V_r^* \gamma_r.
\end{equation}
Taking all these results together, the CorrCal chi-squared can be written as
\begin{equation}
    \chi^2_C = \sum_k \frac{|d_k|^2}{\sigma_k^2} - \sum_r \frac{|V_r|^2 |\gamma_r|^2}{1 + |V_r|^2 \beta_r}.
\end{equation}
If we take the limit $|V_r| \rightarrow \infty$, we get
\begin{equation}
    \lim_{|V_r|\rightarrow\infty} \chi^2_C = \sum_k \frac{|d_k|^2}{\sigma_k^2} - \sum_r \frac{|\gamma_r|^2}{\beta_r} = \chi^2_R \bigg|_{V_r = \hat{V}_r}.
\end{equation}
CorrCal therefore reduces to redundant calibration when we do not include any source modeling, exclude the determinant normalization in the likelihood, and take the limit that the redundant visibilities are large relative to the noise amplitude.

\end{document}